\documentclass[
aps,
prl,
twocolumn
]{revtex4-2}
\usepackage{amsthm}
\usepackage{amsmath,amssymb}
\usepackage{graphicx}
\usepackage{dcolumn}
\usepackage{bm}
\usepackage{xcolor}
\usepackage{siunitx}
\usepackage{hyperref}

\DeclareMathOperator{\Tr}{Tr}

\begin{document}

\title{Realizing Unitary $k$-designs with a Single Quench}

    \author{Yi-Neng Zhou}
    \email{zhouyn.physics@gmail.com}
\affiliation{Department of Theoretical Physics, University of Geneva, 24 quai Ernest-Ansermet, 1211 Genève 4, Suisse}


\author{Robin L\"{o}wenberg}

\affiliation{Department of Theoretical Physics, University of Geneva, 24 quai Ernest-Ansermet, 1211 Genève 4, Suisse}

\author{Julian Sonner}
\email{julian.sonner@unige.ch}
\affiliation{Department of Theoretical Physics, University of Geneva, 24 quai Ernest-Ansermet, 1211 Genève 4, Suisse}

	\begin{abstract}

    We present a single-quench protocol that generates unitary $k$-designs with minimal control. A system first evolves under a random Hamiltonian $H_1$; at a switch time $t_s \geq t_{\mathrm{Th}}$ (the Thouless time), it is quenched to an independently drawn $H_2$ from the same ensemble and then evolves under $H_2$. This single quench breaks residual spectral correlations that prevent strictly time-independent chaotic dynamics from forming higher-order designs. The resulting ensemble approaches a unitary $k$-design using only a single control operation -- far simpler than Brownian schemes with continuously randomized couplings or protocols that apply random quenches at short time intervals. Beyond offering a direct route to Haar-like randomness, the protocol yields an operational, measurement-friendly definition of $t_{\mathrm{Th}}$ and provides a quantitative diagnostic of chaoticity. It further enables symmetry-resolved and open-system extensions, circuit-level single-quench analogs, and immediate applications to randomized measurements, benchmarking, and tomography.

	\end{abstract}

 \maketitle

    {\em Introduction---} Random unitaries play a central role in quantum chaos and thermalization \cite{Roberts_2017, Cotler_2017,Brown_2018,Liu_2018}, quantum computation \cite{bernstein1993quantum}, quantum state tomography \cite{Huang_2020,PhysRevLett.125.200501,Elben_2022,ODonnell2015EfficientQT}, and quantum cryptography \cite{PortmannRenner2022RMP}. In quantum computing, Haar-random unitaries underpin randomized benchmarking \cite{Emerson_2005, Knill_2008} and randomized measurements \cite{PhysRevLett.108.110503,Vermersch_2019,Elben_2019,PhysRevLett.120.050406,Joshi_2020, Elben_2022} and also serve as a resource for demonstrations of quantum-advantage \cite{Arute_2019,Zhong_2020,2022Natur.606...75M}. In many-body dynamics, scrambling and the emergence of Haar-like behavior provide tractable models for chaotic systems and illuminate their spectral statistics and dynamical properties. The Haar measure further enables analytic control through its symmetries and deep connections to random-matrix theory (RMT) \cite{Collins2002,Mehta2004,Collins_2006,Anderson2009,Tao2012}.

Given these broad applications — and because exact sampling from the Haar measure is prohibitively expensive \cite{knill1995approximationquantumcircuits,Barenco1995,Mottonen2005,Mezzadri2006} — a central challenge is to generate ensembles that realize unitary $k$-designs \cite{RandomUnitary2003,dankert2005efficientsimulationrandomquantum,Gross_2007,ambainis2007quantumtdesignstwiseindependence,Roy_2009,PhysRevA.80.012304,Mele_2024}. A unitary $k$-design is an ensemble whose moments match those of the Haar measure up to order $k$ and is a key benchmark of Haar randomness. The goal, therefore, is to realize unitary $k$-designs with minimal experimental control overhead. Much progress has come from circuit implementations, but these either require finely structured gate sequences or rapidly varying random interactions—both of which depart from natural Hamiltonian evolution \cite{RandomUnitary2003, Dankert2009PRA, Brand2016, Brand_o_2016, Brand_o_2021, Schuster:2024ajb, laracuente2024approximateunitarykdesignsshallow,lami2025quantumstatedesignemergent}. By contrast, Hamiltonian dynamics, particularly with time-independent couplings, offer a more practical route for experiments. Also, Floquet driving offers another experimentally relevant route and has been explored extensively \cite{RandomUnitary2003,MerkelRiofrioFlammiaDeutsch2010PRA, HoChoi2022PRL,IppolitiHo2022Quantum, FarshiEtAl2022JMP, quillen2025generatingpseudorandomunitariesfloquet}.

Quantum chaotic systems scramble information: localized information rapidly delocalizes and mixes, suggesting a quantitative link between chaos and Haar randomness \cite{Roberts_2017,Cotler_2017,Brand_o_2016,Harrow_2009,PhysRevX.7.021006,Onorati_2017,Liu_2018,Bhattacharyya_2022,PhysRevX.15.011031,Chenu_2018, Chenu_2019, zhou2026generalizedloschmidtechoinformation}. This motivates using chaotic dynamics to approximate Haar statistics. However, purely time-independent chaotic evolution typically fails to reproduce Haar moments beyond low orders \cite{Roberts_2017, Guo_2024, cui2025randomunitarieshamiltoniandynamics,Chen_2024}. By contrast, Brownian (continuously randomized) evolutions can reach $k$-designs on time/depth poly in $k$ and system size, but at the cost of rapid, sustained modulation of couplings that is difficult to implement experimentally \cite{Jian2022LinearGO, Tiutiakina2023FramePO, Onorati_2017, Brand_o_2021, Guo_2024, Vermersch_2018, tirrito2025anticoncentrationnonstabilizernessspreadingergodic}. These considerations motivate the search for efficient, experimentally viable Hamiltonian protocols that achieve unitary designs with minimal control.

\begin{figure}[t] 
    \centering \includegraphics[width=0.4\textwidth]{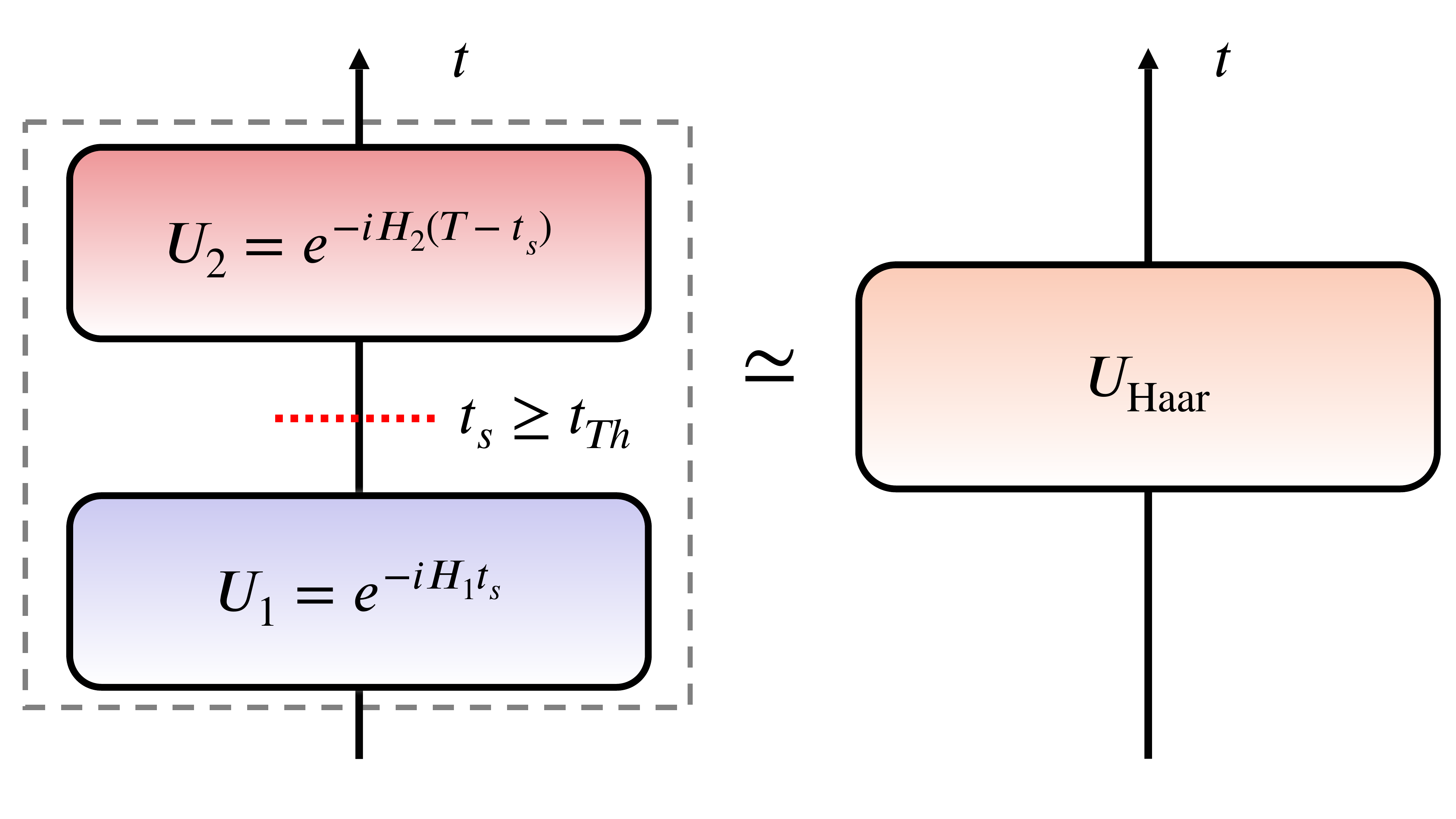} 
    \caption{Single-quench protocol for generating unitary $k$-designs. The system evolves under $H_1$ for $0<t\leq t_s$. At the switch time $t=t_s$, a sudden quench changes the Hamiltonian to $H_2$, under which it evolves for $t_s < t \leq T$. Despite requiring only a single quench step (one switch from $H_1$ to $H_2$), the protocol retains the design-forming power of Brownian schemes and realizes a unitary $k$-design when $t_s\geq t_{\mathrm{Th}}$(the Thouless time).}
\label{single_protocol_fig}
\end{figure}

\begin{figure*}[!t]
  \centering
\includegraphics[width=1.0\textwidth]{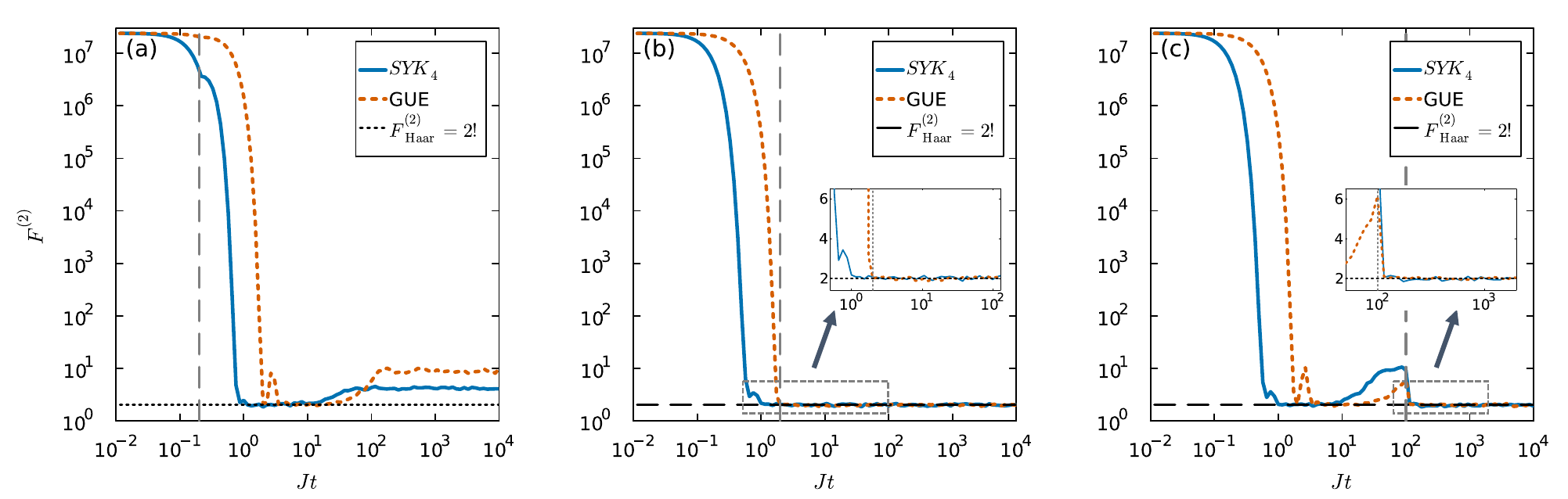}
\caption{
Second-order frame potential $F^{(2)}$ versus total evolution time for single-quench dynamics of the complex-fermion $\mathrm{SYK}_4$ model in a fixed-charge sector (blue solid) and the GUE ensemble (red dashed). SYK parameters: $N=8$, $q=4$ (number of complex fermions), $J=1$; GUE dimension: $N_{\mathrm{dim}}=70$. Curves are averaged over $4000$ disorder realizations. The Haar benchmark $F^{(2)}=2$ is shown as a black dotted line. Gray vertical dashed lines indicate the switch time $t_s$ in each panel: (a) $J t_s=0.2$ (before the Thouless time $t_{\mathrm{Th}}$), (b) $J t_s=2.0$ (near $t_{\mathrm{Th}}$), and (c) $J t_s=100.0$ (after $t_{\mathrm{Th}}$). At late times, $F^{(2)}$ plateaus above the Haar value for $t_s<t_{\mathrm{Th}}$ and reaches the Haar value for $t_s\gtrsim t_{\mathrm{Th}}$. Insets in panels (b) and (c) zoom into the vicinity of $F^{(2)}=2$ near $t\approx t_s$.}
  \label{SYK_GUE_fig}
\end{figure*}

In this Letter, we introduce a single-quench random-Hamiltonian evolution that realizes unitary $k$-designs with minimal control. The system first evolves under a random Hamiltonian $H_1$; at a chosen switch time $t_s$, it is quenched to an independent random Hamiltonian $H_2$ sampled from the same ensemble, and then the system evolves under $H_2$. This single quench disrupts residual spectral and structural correlations that otherwise obstruct design formation, and we show that for $t_s \ge t_{\mathrm{Th}}$ (the Thouless time) the resulting ensemble attains the unitary $k$-design benchmarks. Despite requiring only one control action, the protocol retains the design-forming power of Brownian schemes. We further analyze multi-quench evolution that systematically reduces the design error and achieves a target accuracy $\varepsilon$ with only $O(\log(1/\varepsilon))$ quenches. The approach applies broadly across chaotic platforms (e.g., trapped ions, superconducting qubits, cold atoms), providing a practical route to scalable unitary designs. Finally, the proximity of the single-quench ensemble to Haar randomness yields an operational definition of Thouless time \cite{EdwardsThouless1972,Thouless1974,Beenakker1997,AkkermansMontambaux2007, EversMirlin2008,ChanDeLucaChalker2018} and a quantitative diagnostic of chaoticity.

{\em Single-quench evolution protocol---} 
Consider a system with a random Hamiltonian, e.g., a random matrix sampled from the GUE or a model with random couplings. In the single-quench protocol, the Hamiltonian is
\begin{equation}
H(t)=
\begin{cases}
H_1, & 0\le t<t_s,\\[2pt]
H_2, & t\ge t_s,
\end{cases}
\end{equation}
where $H_1$ and $H_2$ are independent draws from the same random ensemble. In this protocol, we quench the Hamiltonian from $H_1$ to $H_2$ at time $t_s$. The time-evolution operator is therefore
\begin{equation}
U(t;t_s)=
\begin{cases}
e^{-i H_1 t}, & 0\le t\le t_s,\\[6pt]
e^{-i H_2 (t-t_s)}\, e^{-i H_1 t_s}, & t> t_s.
\end{cases}
\end{equation}
Here, $U(t; t_s)$ denotes the total evolution up to time $t$, with a single quench occurring at time $t_s$ \footnote{In practice, our protocol requires (i) two independent realizations of a chaotic Hamiltonian (e.g., two disorder draws of $\mathrm{SYK}_4$) and (ii) evolution for two fixed durations with a single quench at $t_s$. Once $(t_s,\,t-t_s)$ are chosen and calibrated, they are held fixed for all samples; the only per-sample randomness is resampling the disorder couplings defining $(H_1,H_2)$. Time evolution therefore sets the runtime (total evolution time $t$) but adds no extra control parameters, unlike protocols based on many stroboscopic layers or continuously modulated random controls.}. This single-quench evolution protocol is illustrated in Fig.~\ref{single_protocol_fig}.

To certify that the single-quench evolution of chaotic models can realize a unitary $k$-design, we need to quantify the deviation of an ensemble from the Haar random ensemble. This can be done via the frame potential. For an ensemble $\nu\subset U(d)$, its $k$-th order frame potential (FP) \cite{BenedettoFickus2003,RoyScott2007JMP,Scott_2008,CasazzaKutyniok2012,Roberts_2017,Cotler_2017,Mele_2024} is defined as
\begin{equation}
    F^{(k)}_{\nu}\;=\;\mathbb{E}_{U,V\sim \nu}\, \big|\mathrm{Tr}(U^\dagger V)\big|^{2k}, \  k \in \mathbb{N}.
\end{equation}
Among all the random unitary ensembles, the Haar random ensemble minimizes the FP:
\begin{equation}
F^{(k)}_{\nu}\;\ge\;F^{(k)}_{\mathrm{Haar}}=k! \  \  \text{for} \ d \ge k,
\end{equation}
with equality iff $\nu$ is a unitary $k$-design. Thus, the gap
\begin{equation}
\Delta^{(k)}_{\nu}=F^{(k)}_{\nu}-F^{(k)}_{\mathrm{Haar}}
\end{equation}
provides a convenient scalar measure of “distance” to Haar; $\Delta^{(k)}=0$ certifies a unitary $k$-design.

{\em Single-quench chaotic evolution realizes unitary design---} We simulate single-quench evolution in two chaotic ensembles: the complex-fermion Sachdev--Ye--Kitaev (SYK) model~\cite{SachdevYe1993PRL,Kitaev2015Talks,PolchinskiRosenhaus2016JHEP,MaldacenaStanford2016PRD,MaldacenaStanfordYang2016PTEP,KitaevSuh2018JHEP,BagretsAltlandKamenev2016NPB,BagretsAltlandKamenev2017NPB,JevickiSuzukiYoon2016JHEP,GarciaGarciaVerbaarschot2016PRD,GarciaGarciaVerbaarschot2017PRD,GuQiStanford2017JHEP,SongJianBalents2017PRL,Rosenhaus2019JPA,ChowdhuryGeorgesParcolletSachdev2022RMP} (blue solid curves in Fig.~\ref{SYK_GUE_fig}) and the Gaussian unitary ensemble (GUE) of random matrices (red dotted curves in Fig.~\ref{SYK_GUE_fig}). For each sample, we draw two independent Hermitian Hamiltonians $(H_1, H_2)$ from the chosen ensemble (SYK or GUE), generate the single-quench unitary $U(t;t_s)$, and evaluate the FP as a function of the total evolution time $t$. For the SYK$_4$ model, the system consists of complex fermions with all-to-all interactions, and its Hamiltonian is given by
\begin{equation}
    \label{eqn:SYK4}
    H_{\text{SYK}_4} = \sum_{ 1\leq i<j<k<l \leq N}J_{ijkl} c_i^{\dagger} c_j^{\dagger} c_k c_l,
\end{equation}
where  $c_i(c_i^{\dagger})$ denotes the annihilation(creation) operator of complex fermions that satisfies the anticommutation relation $\{c_i,c_j^{\dagger}\} = \delta_{ij}$. Here, $J_{ijkl}$ is a Gaussian random variable with zero mean and variance 
\begin{equation*} 
	\langle J_{ijkl} J_{i^{'}j^{'}k^{'}l^{'}} \rangle=\delta_{i,i^{'}}\delta_{j,j^{'}}\delta_{k,k^{'}}\delta_{l,l^{'}}\frac{3!J^2}{N^{3}}.
\end{equation*}
Experimental proposals to realize SYK models span Majorana devices, ultracold atoms, graphene in strong fields, and cavity QED~\cite{PikulinFranz2017PRX,DanshitaHanadaTezuka2017PTEP,ChenIlanDeJuanPikulinFranz2018PRL,UhrichBandyopadhyaySauerweinSonnerBrantutHauke2023,Cao_2020, baumgartner2024quantumsimulationsachdevyekitaevmodel}, highlighting both the experimental push toward SYK models and the utility of low-control protocols with clear scrambling and near-Haar signatures.

In both cases, when the switch time $t_s$ is at or beyond the Thouless time $t_{\mathrm{Th}}$, its $k$-th FP saturates the Haar value, indicating that the protocol realizes a unitary $k$-design (see Fig.~\ref{SYK_GUE_fig}(b),(c) for $k=2$) \footnote{A chaotic model for unitary design is needed because it must rapidly scramble, so that for times $t \gtrsim t_{\mathrm{Th}}$ the evolution exhibits universal RMT features. Equally important, the quench must \emph{inject independence}: changes that act mainly by conjugation are mere ``rotations'' that preserve the spectrum, whereas switching to an independent chaotic Hamiltonian generically changes both eigenvalues and eigenvectors.}. For $t_s<t_{\mathrm{Th}}$, the late-time plateau lies above the Haar value (see Fig.~\ref{SYK_GUE_fig}(a)), so a unitary $k$-design is not achieved. While Fig.~\ref{SYK_GUE_fig}(c) shows that both GUE and SYK can transiently approach the $k$-design value near the $t_{\mathrm{Th}}$, this occurs only in a narrow time window and would require fine-tuned sampling. By contrast, a single quench applied after $t_{\mathrm{Th}}$ yields a robust late-time plateau, and it continues to work for less chaotic or modified SYK variants where the no-quench dynamics do not achieve comparable design quality (see supplementary material (SM) \cite{SM}).

\begin{figure}[t] 
    \centering \includegraphics[width=0.4\textwidth]{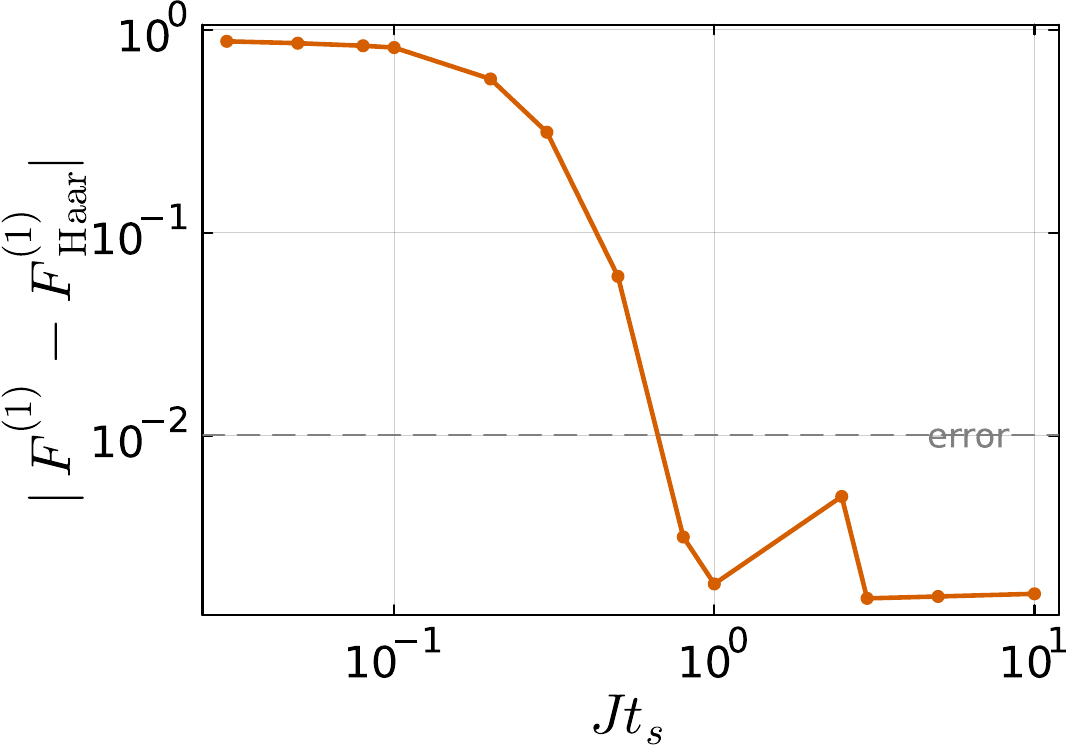} 
    \caption{Gap $\Delta_{\mathrm{FP}}(t_s)$ between the late-time plateau of the first-order frame potential $F^{(1)}$ and its Haar value for single-quench evolution of the complex SYK$_4$ model in a fixed charge sector, plotted versus the switch time $t_s$. Parameters: $N=6$, total charge $q=2$, $J=1$; averaged over $N_{\mathrm{rand}}=10^4$ disorder realizations. The Thouless time $t_{\mathrm{Th}}$ is defined as the first $t_s$ at which $\Delta_{\mathrm{FP}}(t_s)$ is consistent with zero within the statistical uncertainty (gray dashed line). The uncertainty is estimated as $\sqrt{1/[\sqrt{N_{\mathrm{rand}}}(\sqrt{N_{\mathrm{rand}}}-1)]}\simeq 0.01$ \cite{SM}.}
\label{CSYK4_crossover_fig}
\end{figure}

As $t_s$ approaches $t_{\mathrm{Th}}$ from below, the gap $\Delta_{\mathrm{FP}}(t_s)$ between the late-time plateau and the Haar value decreases monotonically. Plotting $\Delta_{\mathrm{FP}}(t_s)$ versus $t_s$ shows that it approaches zero and then remains small and nearly flat for $t_s \gtrsim t_{\mathrm{Th}}$, as illustrated in Fig.~\ref{CSYK4_crossover_fig} for the single-quench time evolution of the complex SYK model. See the error bar calculation in SM \cite{SM}. This behavior motivates an operational definition of the Thouless time:
\begin{equation}
\begin{aligned}
t_{\mathrm{Th}}
&:= \min\bigl\{\, t_s : \Delta_{\mathrm{FP}}(t_s) \text{ decreases to zero}\\
&\qquad\text{(within statistical uncertainty)} \bigr\}.
\end{aligned}
\end{equation}
Previous works often estimate $t_{\mathrm{Th}}$ from the onset of the linear ramp in the spectral form factor (SFF) or from related survival-probability diagnostics~\cite{Gharibyan:2018jrp, Schiulaz_2019}.  
At finite size, however, the ramp crossover is strongly oscillatory, making the inferred onset sensitive to smoothing/fitting choices. By contrast, for the quench FP we show analytically (for GUE in Eq.\eqref{eq_GUE_FP} that the deviation from the Haar value is controlled by a term $\propto \mathrm{SFF}^2$, whose decaying envelope suppresses these oscillations.  
This yields a smoother approach to the Haar value and supports a more robust operational definition of $t_{\mathrm{Th}}$, e.g.\ as the earliest time at which the FP stays within a prescribed tolerance of its Haar value (see SM for details).  

\begin{figure*}[t]
  \centering
\includegraphics[width=0.85\textwidth]{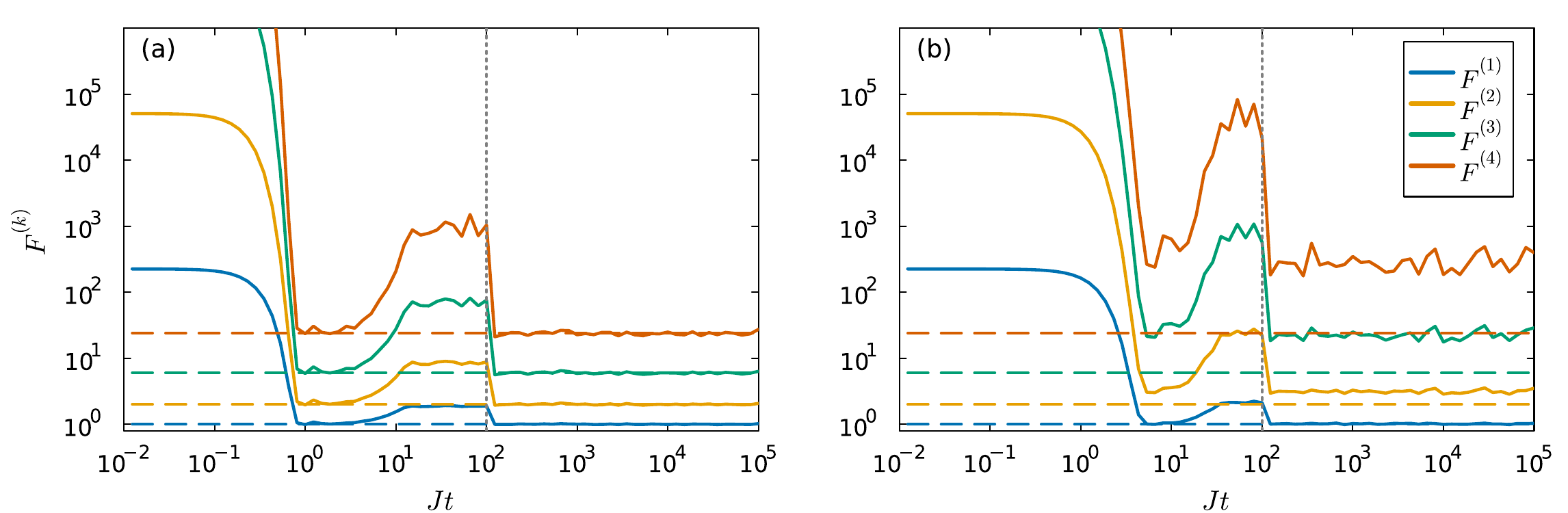}
  \caption{Frame potentials $F^{(k)}$ for orders $k=1,2,3,4$ versus total evolution time $Jt$ under the single-quench evolution in (a) complex SYK$_4$ model and (b) complex SYK$_2$ model. Curves for $k=1,2,3,4$ are shown in blue, orange, green, and vermilion, respectively; the Haar benchmarks $F_{\mathrm{Haar}}^{(k)}=k!$ appear as dashed lines of matching colors. The gray dotted vertical line marks the switch time $Jt_s=100$. Parameters: $N=6$ complex fermions in fixed-charge sector $q=2$, $J=1$; averages over $10^4$ disorder realizations. In (a), SYK$_4$ saturates the Haar values up to $k=4$, whereas in (b), SYK$_2$ achieves only the $k=1$ benchmark.}
  \label{fig:high_k_compare}
\end{figure*}

The single-quench protocol reaches a unitary design on the Thouless-time scale of the underlying dynamics. In strongly chaotic models, one expects this Thouless time to be as short as possible. For example, the SYK model has been argued to have a Thouless time of order $\sqrt{N} \log N$ in a supersymmetric sigma model approach \cite{Altland:2017eao,Altland:2021rqn}. It would be interesting to investigate whether a certain class of models might exhibit a fast-ergodic behavior\footnote{The model has been argued to have a Thouless time of order $\log N$ in \cite{Gharibyan:2018jrp}.}, analogous to the fast-scrambling behavior\cite{Sekino_2008}. In contrast, for an integrable random Hamiltonian—such as the complex SYK$_2$ model—the same single-quench protocol fails to generate higher-order unitary $k$-designs. Thus, realizability within this protocol could serve as an operational diagnostic of quantum chaos. Figure~\ref{fig:high_k_compare} plots the FP $F^{(k)}$ for $k=1,2,3,4$ versus total evolution time $t$ for (a) the chaotic complex SYK$_4$ model and (b) the integrable complex SYK$_2$ model. In SYK$_4$, $F^{(k)}$ saturates the Haar benchmarks $F_{\mathrm{Haar}}^{(k)}=k!$ up to $k=4$, demonstrating unitary $k$-design behavior through fourth order, whereas SYK$_2$ reaches only the $k=1$ (unitary 1-design) value \footnote{Although the complex SYK$_2$ model does not realize a $k$-design, it still reaches the Brownian value also for higher orders 
$k$; see SM \cite{SM} for details.}. Additional results, including SYK$_2$ and the Richardson model (both interacting and integrable), are presented in SM \cite{SM}.

   {\em Random matrix analysis.---} We next analyze the single-quench evolution with Hamiltonians sampled from the GUE using RMT analysis, focusing on the first-order FP as a simple, explicit example \footnote{For SYK, an analytic evaluation of the single-quench FP is technically challenging and, to our knowledge, no closed-form result is available even in closely related settings. We therefore rely on numerics for SYK, while using our GUE analytics as guidance: beyond the Thouless time, SYK exhibits random-matrix spectral correlations, so the GUE frame-potential calculation captures the expected late-time behavior and provides intuition for the approach to an approximate $k$-design.}. We will show that the term controlling the FP-Haar deviation vanishes in the thermodynamic limit once the switch time exceeds the Thouless time. This term is expressed directly via the SFF, making the “FP-Haar gap” definition of the Thouless time equivalent to the standard definition based on the onset of the SFF linear ramp.

For reference, the first-order FP for the unitary time evolution generated from GUE is written as
\begin{equation}
F^{(1)}(t)
= \int dH_1\,dH_2\, P(H_1)P(H_2)\,\big|\mathrm{Tr}\big(e^{iH_1 t}e^{-iH_2 t}\big)\big|^2.  
\end{equation}
Here, $P(H_i) = e^{-\frac{d}{2}\text{Tr}(H_i^2)}$. For the single-quench of total evolution time $t$ and switch $H$ at time $t_s$, its $1$-st FP is 
\begin{equation}
\begin{aligned}
&F^{(1)}(t;t_s)=  \int \Big( \prod_{a=1}^4   dH_a P(H_a) \Big) \quad \\
&\times
\big|\mathrm{Tr}\left(e^{iH_1(t-t_s)}e^{iH_2 t_s}e^{-iH_3 t_s}e^{-iH_4(t-t_s)}\right)\big|^2,
\end{aligned}
\end{equation}
with $H_1,H_2,H_3,H_4$ i.i.d. GUE.

Write $e^{iH_a t}=V_a \Lambda^{a} V_a^\dagger$ ($\Lambda^a$ is diagonal), and use the unitary-conjugation invariance of the GUE measure. Defining $U^4:=(U^1U^2U^3)^\dagger$ and $U^a:=V_a^\dagger V_{a+1} \ (a=1,2,3)$, one obtains
\begin{equation}
\begin{aligned}
    F^{(1)}(t;t_s)
    = &  \int \Big( \prod_{a=1}^4 D\lambda_a \Big)\int_{\mathrm{Haar}} dU^1 dU^2 dU^3\\
&\Big|\mathrm{Tr}\left(\Lambda^{1\dagger}U^1 \Lambda^{2\dagger}U^2 \Lambda^{3} U^3 \Lambda^{4} U^4\right)\Big|^2 .
\end{aligned} 
\end{equation}
Performing the Haar integrals over $U^{1},U^{2},U^{3}$ using second-moment Weingarten calculus, the leading contribution at large Hilbert-space dimension $d$ is
\begin{equation}
F^{(1)}(t;t_s)
= 1 +d^2\left(\frac{R_2(t-t_s)}{d^2}\right)^{2}
  \left(\frac{R_2(t_s)}{d^2}\right)^{2}
+ \mathcal{O}\left(d^{-1}\right).
  \label{eq_GUE_FP}
\end{equation}
Here, $R_2(t)$ is the \textit{spectral form factor} (SFF) defined as
\begin{equation}
R_2(t)=\int D\lambda \sum_{i,j} e^{,i(\lambda_i-\lambda_j)t}. 
\label{SFF_definition}
\end{equation}
For GUE, one may write
\begin{equation}
R_2(t)= d + d^2 r_1^2(t) + d \ r_2(t),
\end{equation}
where $r_1(t)$ and $r_2(t)$ encode the familiar dip-ramp-plateau structure \cite{Cotler_2017} (their explicit forms are not needed below). Equation \eqref{eq_GUE_FP} reduces to the single evolution (without single quench) result at $t_s=0$. At the dip (Thouless) time $t_{\mathrm{Th}}$, $R_2(t_s)\sim \sqrt{d}$; on the plateau, $R_2(t_s)\sim d$. Hence, if the switch time satisfies $t_s\ge t_{\mathrm{Th}}$, the second term in \eqref{eq_GUE_FP} vanishes in the thermodynamic limit $d\to\infty$. Consequently,
$F^{(1)}(t;t_s)\to 1$,
the Haar value of the first-order FP. 

Intuitively, replacing a single uninterrupted evolution with a single quench factorizes the deviation from the Haar average into two SFF-controlled factors, one from the segment of length $t_s$ and one from the segment of length $t-t_s$. The SFF contribution from the first segment decays from $t=0$ to the dip and becomes parametrically small once $t_s \gtrsim t_{\mathrm{Th}}$; multiplying by this small factor suppresses the overall deviation to $\mathcal{O}(1/d^2)$, which vanishes as $d \to \infty$. Without the quench, this extra suppression is absent, so the long-time deviation remains set by the late-time SFF and is $\mathcal{O}(1)$. In particular, the deviation from the unitary 1-design value is bounded by $\mathcal{O}(1/d^2)$: in the large-$d$ limit, this single-quench protocol realizes an exact unitary 1-design, while for finite $d$ it achieves an approximate 1-design with relative error $\mathcal{O}(1/d^2)$. Multiple quenches further reduce this error.

The same structure persists for higher-order FP: the deviation from the Haar value is a product of higher-order SFF factors and vanishes in the thermodynamic limit once $t_s>t_{\mathrm{Th}}$. Therefore, locating $t_{\mathrm{Th}}$ via the closing of the FP-Haar gap is equivalent to identifying the onset of the SFF’s linear ramp, providing an operational and experimentally convenient definition of the Thouless time.

One may ask whether the convergence to a $k$-design could be improved even further by re-sampling $m$ times during the protocol. For  an quench $m$-time ($m>1$) FP, the leading order contribution is 
\begin{equation*}
\begin{split}
    F_m^{(1)}(t_1,t_2,...,t_{m+1})=1+ d^2\prod_{j=1}^{m+1}\left(\frac{R_2(t_j)}{d^2}\right)^2 + \mathcal{O}\left(d^{-1}\right).
\end{split}
\end{equation*}
Here, $F_m^{(1)}(t_1, t_2, \ldots, t_{m+1})$ denotes the case where the $j$-th quench is performed at time $t = \sum_{r=1}^{j} t_r$. If we assume each time segment is the same, then we have 
\begin{equation*}
    F^{(1)}_m(\Delta t,\dots,\Delta t)=1+ d^2\left(\frac{R_2(\Delta t)}{d^2}\right)^{2(m+1)} + \mathcal{O}\left(d^{-1}\right).
    \label{Eq_quench-m-time}
\end{equation*}
Thus, by fixing a small time step $\Delta t$ and evaluating its SFF, we can compute the corresponding correction using Eq.~\eqref{eq_GUE_FP}. If we aim to realize a unitary $1$-design with error smaller than $\epsilon$, above Equation implies that the quench time must satisfy
$m > \frac{\log\!\bigl(\frac{d^{2}}{\epsilon}\bigr)}{2\,\log\!\left[\frac{d^{2}}{R_{2}(\Delta t)}\right]} - 1$.

{\em Approximate $k$-designs via multiple-quench---} 
In realistic settings, one only needs to reach a unitary $k$-design up to a relative error $\epsilon$. We say an ensemble $\nu$ on $U(d)$ is an Tensor Product Expander (TPE) $\epsilon$-approximate unitary $k$-design \cite{HastingsHarrow2009QIC,Dankert2009PRA,HarrowLow2009ApproxRandom,Brand_o_2016,Sen2018ZigzagTPE,GrossAudenaertEisert2007JMP,Haferkamp2022Quantum} if and only if 
\begin{equation}
     \Bigl\| \mathbb{E}_{V\sim\varepsilon}\!\big[\,V^{\otimes k}\otimes V^{*\otimes k}\,\big]-\mathbb{E}_{U\sim \mu_{H}}\!\big[\,U^{\otimes k}\otimes U^{*\otimes k}\,\big] \Bigr\|_{\infty}\leq \epsilon.
\end{equation}
Using the amplification property of TPE ~\cite{Haferkamp2022Quantum}: 
for $1 \ge \epsilon_{0} \ge \epsilon > 0$, if a unitary ensemble $\nu$ is a TPE
$\epsilon_{0}$-approximate $k$-design, then the ensemble $\nu_{P}$ of products
\begin{equation*}
    U_{P} \cdots U_{1}, \qquad U_{j} \sim \nu \text{ i.i.d.},
\end{equation*}
is an $\epsilon$-approximate $k$-design where $P$ is a positive integer such that $P\geq \frac{\log(1/\epsilon)}{\log(1/\epsilon_0)} $.
Thus, one can use the multiple-quench protocol to reach an approximate $k$-design. For a single-quench evolution of total time $\Delta t$ that reaches a $\epsilon_0$-approximate $k$-design, one can quench the evolution every $\Delta t$ interval for $M$ time to achieve an $\epsilon$-approximate $k$-design, and we only need $M\ \ge\ \frac{\log(1/\epsilon)}{\log(1/\epsilon_0)}-1.$
The number of quench and also the total evolution time, which is $T=(M+1)\Delta t$, grow only logarithmically with the target accuracy, a favorable scaling for experiments. A proof of this error amplification for unitary $k$-designs is provided in the SM \cite{SM}.

{\em Conclusion and discussion---} In this Letter, we show that a \emph{single-quench} evolution of a quantum-chaotic system can realize unitary $k$-designs, supported by both analytical results (for GUE) and numerical frame-potential estimates (for SYK and other models). Leveraging fast scrambling, chaotic dynamics serve as efficient randomness generators: unlike strictly time-independent chaotic Hamiltonians—which typically fail beyond low moments—the single-quench protocol attains Haar-level frame potentials once the switch time $t_s$ reaches the Thouless time $t_{\mathrm{Th}}$. In contrast to Brownian (continuously randomized) schemes that require rapid, persistent modulation of couplings, our protocol uses only one quench, greatly simplifying implementation\footnote{We illustrate this with the complex SYK model and the GUE random-matrix model; the protocol also extends to other systems, such as the Majorana SYK model (see \cite{SM}).}. Moreover, in the more realistic setting where we aim to realize approximate unitary $k$-designs with relative error $\epsilon$, a multiple-quench protocol can be used to reduce the error. The required number of quenches is proportional to 
$\log{(1/\epsilon)}$. Beyond providing a simple route to Haar-like randomness, the protocol yields an \textit{operational} definition of the Thouless time: $t_{\mathrm{Th}}$ is the smallest switch time $t_s$ after which the late-time frame-potential gap to Haar vanishes within statistical uncertainty. This agrees with the spectral criterion (onset of the ramp in the SFF) while offering an experiment-friendly measurement procedure. Moreover, the single-quench realizability furnishes a quantitative diagnostic of the underlying model’s chaoticity.

Our work opens several concrete directions: $(\mathrm{i})$ \emph{Integrability and weak chaos:} How effective is the single-quench protocol in the integrable or near-integrable regime? $(\mathrm{ii})$ \emph{Minimal perturbations:} Replacing the quench with a weak random perturbation may provide an even lighter-weight route to unitary designs. $(\mathrm{iii})$\emph{Symmetries and sectors:} With conservation laws or fixed-sector dynamics \cite{Jian2022LinearGO, Tiutiakina2023FramePO}, how do symmetry constraints modify the attainable design order and the required switch time $t_s$? (iv) \emph{Transition-like behavior:} Varying the ratio of random to deterministic couplings could induce sharp crossovers in “designability,” potentially correlated with underlying dynamical or spectral transitions. $(\mathrm{v})$ \emph{Experimental integration:} The protocol is naturally compatible with randomized measurements (OTOCs, entanglement probes), since the Hamiltonian already realized on hardware can generate near-Haar unitaries after a single parameter switch. Identifying optimal operating points would directly benefit current platforms. $(\mathrm{vi})$ \emph{Beyond Hamiltonian:} Our single-quench protocol suggests circuit-level analogs: a one-shot reconfiguration of the gate ensemble that approaches Haar moments with minimal spatiotemporal modulation, providing an alternative to fully Brownian schemes. A key question is how the condition $t_s \gtrsim t_{\mathrm{Th}}$ maps onto a depth (or light-cone) requirement in random-circuit realizations \footnote{Our protocol is intrinsically \emph{analog} (quench evolution under a many-body Hamiltonian), so a gate \emph{depth} is not an inherent notion. 
For digital implementations, a Trotterized realization of full SYK would require a depth growing with system size and the number of interaction terms. 
By contrast, in the reduced-rank SYK setting, the same $k$-design behavior can be reached with effectively constant control complexity: in the proposal \cite{baumgartner2024quantumsimulationsachdevyekitaevmodel}, the random couplings are imprinted by a single speckle pattern, and a quench corresponds to one rapid reconfiguration, i.e.\ two global unitaries (before/after the quench) independent of system size.}. Control-parameter complexity is also an interesting open question \footnote{An open question is how many independent couplings must be re-randomized across a quench to achieve an approximate $k$-design, and whether a small set of control knobs already suffices. Clarifying this would connect our analog sampler to random-circuit and Floquet-based $k$-design constructions, where the resources are naturally expressed in terms of depth and tunable parameters.} $(\mathrm{vii})$  \emph{Lindbladian evolution:} Does the single-quench idea extend to open, Markovian dynamics? Can one quench dissipative rates or the Lindbladian generator to realize approximate unitary $k$-designs—or channel $k$-designs over CPTP maps? $(\mathrm{viii})$  \emph{Strong designs:} Can single-quench evolution realize strong unitary designs \cite{PRXQuantum.2.030339, Bannai2022, schuster2025strongrandomunitariesfast}? Overall, our single-quench approach provides a minimal-control pathway to high-quality pseudorandomness. Mapping its performance across symmetry classes, degrees of integrability breaking, and architectures—and establishing rigorous finite-size and finite-time error bounds for approximate $k$-designs—are natural next steps with clear theoretical and experimental payoff.

 {\em Acknowledgments---} We thank Thierry Giamarchi, Romain Vasseur, Laura Foini, Chang Liu, Tian-Gang Zhou, Pengfei Zhang, and Liang Mao for helpful discussions. This work has received support through the Swiss Quantum Initiative awarded by the State Secretariat for Economic Affairs, under the grant "HoloGraph". This research is supported in part by the Fonds National Suisse de la Recherche Scientifique (Schweizerischer Nationalfonds zur Förderung der wissenschaftlichen Forschung) through the Project Grant 200021\_215300 and the NCCR51NF40-141869 The Mathematics of Physics (SwissMAP).

 \appendix

 \newpage

\onecolumngrid

\appendix
\section*{Supplementary Material}
\addcontentsline{toc}{section}{Supplementary Material}

In this supplementary material, we present:
(A) a detailed computation of the frame potential (FP) for the multiple-quench protocol; (B) a step-by-step derivation of the first-order FP for single-quench evolution generated by GUE Hamiltonians; (C) additional numerical results for the single-quench FP in integrable models (the free SYK$_2$ and the interacting Richardson model); (D) additional numerical results for single-quench dynamics in chaotic models; (E) a brief overview of analytic results for the complex Brownian SYK model and (F) the error-bar calculation for the finite-sample estimate of the $k$-th FP used in the main-text Fig 3.

\section{Appendix A: Computation of the error of FP from single-quench to multiple-quench}\label{appendix_switch_two_time_error}
 Although the single-quench protocols in the main text already produce excellent approximate $k$-designs in sufficiently chaotic regimes, experimental constraints may limit either the accessible evolution time or the degree of chaos. In such cases, a single quench reaches only a finite accuracy, which we quantify by an error $\epsilon_0$. A standard way to improve the design quality is then to \emph{concatenate} quenches: repeated composition contracts the deviation from Haar, so that after $M$ quenches the error is amplified down to a target $\epsilon$, requiring only $M=O(\log(1/\epsilon))$ as long as $\epsilon_0<1$.

This multiple-quench protocol is an instance of a well-known amplification mechanism: composing unitaries drawn from an $\epsilon_0$-approximate $t$-design (or, more generally, a tensor-product expander) typically improves the design quality under iteration. Rigorous formulations of this contraction have been developed in the context of local random circuits as approximate polynomial designs~\cite{Brand_o_2016} and in the TPE/$t$-design amplification arguments reviewed in Ref.~\cite{Mele_2024}. Below, we adapt these general principles to bound the number of quenches needed to reach a desired accuracy in our setting.

In this appendix, we compute the $k$-th frame potential (FP) for the multiple-quench evolution protocol. Since the Hilbert space dimension is finite in realistic settings, we quantify convergence to Haar up to a tolerance $\epsilon$. Specifically, we say that an ensemble $\varepsilon$ on $U(d)$ forms a $\delta$-approximate unitary $k$-design if its $k$-FP satisfies
\begin{equation}
F^{(k)}_{\varepsilon}-F^{(k)}_{\mathrm{Haar}} \;\le\; \delta.
\end{equation}

For an ensemble $\varepsilon\subset U(d)$, define its $(k,k)$ moment operator

$$
\Phi\equiv \Phi^{(k)}_\varepsilon\;=\;\mathbb{E}_{U\sim\varepsilon}\!\big[\,U^{\otimes k}\otimes U^{*\otimes k}\,\big],
$$

which is Hermitian, positive semidefinite, with $\|\Phi\|_{\mathrm{op}}\le 1$.
Let $\Pi$ denote the Haar moment operator; $\Pi$ is a projector with

$$
\Pi^2=\Pi,\qquad 
\mathrm{Tr}\,\Pi=\mathrm{Tr}\,\Pi^2=F^{(k)}_{\mathrm{Haar}}=
\begin{cases}
k!, & d\ge k,\\[2pt]
\displaystyle \sum_{\lambda\vdash k,\ \ell(\lambda)\le d}(f^\lambda)^2, & d<k.
\end{cases}
$$

The $k$-th FP is

$$
F^{(k)}_{\varepsilon}
=\mathbb{E}_{U,V\sim\varepsilon}\big|\mathrm{Tr}(U^\dagger V)\big|^{2k}
=\mathrm{Tr}(\Phi^2).
$$

Assume the single-quench protocol produces an ensemble $\varepsilon$ with
$$
F^{(k)}_{\varepsilon}=F^{(k)}_{\mathrm{Haar}}+\delta
\quad\Longleftrightarrow\quad
\mathrm{Tr}\big[(\Phi-\Pi+\Pi)^2\big]=F^{(k)}_{\mathrm{Haar}}+\delta.
$$

Work in a basis splitting the Haar subspace and its orthogonal complement:

$$
\Phi\;=\;\Pi\ \oplus\ \operatorname{diag}(\mu_1,\mu_2,\ldots),
\qquad 0\le \mu_i\le 1.
$$

Then

$$
F^{(k)}_{\varepsilon}=F^{(k)}_{\mathrm{Haar}}+\sum_i \mu_i^2
\quad\Rightarrow\quad
\sum_i \mu_i^2=\delta.
$$

For the quench-$3$-time statistic, set $A=U_1U_2$, $B=U_3U_4$ with i.i.d. draws $U_j\sim\varepsilon$. Then $A,B$ are independent with distribution $\varepsilon^{*2}$ (free multiplicative convolution), hence

$$
\mathbb{E}_{U_1,\ldots,U_4\sim\varepsilon}\big|\mathrm{Tr}(U_1U_2U_3^\dagger U_4^\dagger)\big|^{2k}
=F^{(k)}_{\varepsilon^{*2}}.
$$

Under convolution, moment operators multiply:

$$
\Phi(\varepsilon^{*2})=\Phi(\varepsilon)^2=\Phi^2,
$$

so

$$
F^{(k)}_{\varepsilon^{*2}}
=\mathrm{Tr}\!\big(\Phi(\varepsilon^{*2})^2\big)
=\mathrm{Tr}(\Phi^4).
$$

In the same eigenbasis,

$$
\Phi^4=\Pi\ \oplus\ \operatorname{diag}(\mu_1^4,\mu_2^4,\ldots),
$$

and we obtain the exact identity

\begin{equation}
\mathbb{E}\big|\mathrm{Tr}(U_1U_2U_3^\dagger U_4^\dagger)\big|^{2k}
=F^{(k)}_{\mathrm{Haar}}+\sum_i \mu_i^{4}  
\end{equation}
with $\sum_i \mu_i^2=\delta$. Because $0\le \mu_i^4\le \mu_i^2$ and by Cauchy-Schwarz $\sum_i \mu_i^4 \le (\sum_i \mu_i^2)^2=\delta^2$,

\begin{equation}
F^{(k)}_{\mathrm{Haar}}\ \le\ F^{(k)}_{\varepsilon^{*2}}\ \le\ F^{(k)}_{\mathrm{Haar}}+\min\{\delta,\ \delta^2\}. 
\end{equation}

Thus, the FP excess above Haar cannot increase under one convolution step, and when $\delta\ll 1$ it improves quadratically: the four-draw statistic is closer to $F^{(k)}_{\mathrm{Haar}}$ than the two-draw one.

Consider then using iterating quenches to reach an approximate $k$-design. After $M= 2^m-1$ independent quenches (i.e., $2^m$ draws and $2^{m-1}$ convolutions),

$$
\Phi\ \longmapsto\ \Phi^{2^{m-1}},\qquad
F_k(\varepsilon^{*2^{m-1}})=\mathrm{Tr}(\Phi^{2^{m-1}})
=F^{(k)}_{\mathrm{Haar}}+\sum_i \mu_i^{\,2^{m-1}}.
$$

Hence for small $\delta$,

$$
0\le F_k(\varepsilon^{*(2^{m-1})})-F^{(k)}_{\mathrm{Haar}}
\le \Big(\sum_i \mu_i^2\Big)^{2^{m-1}}=\delta^{\,2^{m-1}}.
$$

To achieve an $\epsilon$-approximate $k$-design it suffices that $\delta^{2^{m-1}}\le \epsilon$, i.e.
\begin{equation}
M\ \ge\ 2 \frac{\log(1/\epsilon)}{\log(1/\delta)} -1.
\end{equation}

Thus, the number of quenches grows only logarithmically with the target accuracy (for fixed initial gap $\delta$), a favorable scaling for experiments.

\paragraph{Relation between moment-operator and frame-potential criteria.}
The two error notions used in this work are closely related. 
Let $\mathcal{M}_\varepsilon^{(k)}$ and $\mathcal{M}_{\mathrm{Haar}}^{(k)}$ denote the $k$-th moment operators (twirling channels) of the ensemble $\varepsilon$ and of the Haar measure, respectively. 
The frame potential can be written as the Hilbert–Schmidt norm of the difference of moment operators,
\begin{equation}
F^{(k)}_\varepsilon - F^{(k)}_{\mathrm{Haar}}
= \bigl\| \mathcal{M}_\varepsilon^{(k)} - \mathcal{M}_{\mathrm{Haar}}^{(k)} \bigr\|_2^2 ,
\end{equation}
so that a small frame-potential deviation directly implies that the two moment operators are close in Hilbert–Schmidt norm. 
Since for finite-dimensional spaces the operator norm is bounded by the Hilbert–Schmidt norm,
\(
\|\cdot\|_\infty \le \|\cdot\|_2,
\)
a bound on the frame potential also yields a bound in operator norm (up to dimension-dependent constants, if one compares normalized channels). 
Thus the frame-potential criterion and the moment-operator criterion quantify the same notion of proximity to a $k$-design, differing only by the choice of norm and corresponding prefactors.

\section{Appendix B: Computation of the frame potential of single-quench GUE evolution}
\label{appendix_GUE}

This appendix provides a step-by-step derivation of the first-order frame potential for a single-quench evolution generated by Gaussian Unitary Ensemble (GUE) Hamiltonians.

\paragraph*{Definition.}
For a single, uninterrupted (no quench) evolution, the first-order frame potential is
\begin{equation}
F^{(1)}(t)
=\!\!\int\! dH_1\, dH_2\, P(H_1)P(H_2)\,\bigl|\mathrm{Tr}\!\left(e^{i H_1 t}e^{-i H_2 t}\right)\bigr|^2 .
\end{equation}
For a single quench at time $t_s\in[0,t]$, we write the two segments explicitly and duplicate the Hamiltonians for the complex conjugate trace:
\begin{equation}
\begin{aligned}
F^{(1)}(t;t_s)
&=\int \Big(\prod_{a=1}^4 dH_a P(H_a) \Big) \;
\Bigl|\mathrm{Tr}\!\Big(e^{i H_1 (t-t_s)} e^{i H_2 t_s} e^{-i H_3 t_s} e^{-i H_4 (t-t_s)}\Big)\Bigr|^2 .
\end{aligned}
\label{eq:gue_single_quench_def}
\end{equation}
Here $P(H_j) = e^{-\frac{d}{2}\text{Tr}(H_j^2)}$ is the GUE measure and $d$ denotes the Hilbert-space dimension.

\paragraph*{Diagonalization and unitary invariance.}
Use the spectral decompositions
\begin{equation}
e^{i H_a \tau_a} = V_a \Lambda^{(a)} V_a^\dagger,\qquad
\Lambda^{(a)}=\mathrm{diag}\!\big(e^{i \lambda^{(a)}_1 \tau_a},\dots,e^{i \lambda^{(a)}_d \tau_a}\big),
\end{equation}
with segment durations $(\tau_1,\tau_2,\tau_3,\tau_4)=(t-t_s,\, t_s,\,-t_s,\,-(t-t_s))$.
By unitary-conjugation invariance of GUE, the joint distribution factorizes into eigenvalues and independent Haar unitaries for the eigenvectors. Defining
\begin{equation}
U^1 = V_1^\dagger V_2,\quad
U^2 = V_2^\dagger V_3,\quad
U^3 = V_3^\dagger V_4,\quad
U^4 = (U^1 U^2 U^3)^\dagger,
\end{equation}
and denoting $D\lambda_a$ the induced eigenvalue measures, the trace becomes
\begin{equation}
\begin{aligned}
F^{(1)}(t;t_s)
&=\!\!\int \!\Big(\prod_{a=1}^4 D\lambda_a\Big)\!
\int_{\mathrm{Haar}} dU^1 dU^2 dU^3\;
\mathrm{Tr}\!\big(\Lambda^{(1)\!\dagger} U^1 \Lambda^{(2)\!\dagger} U^2 \Lambda^{(3)} U^3 \Lambda^{(4)} U^4 \big)\\
&\hspace{5.7em}\times
\mathrm{Tr}\!\big(U^{4\!\dagger} \Lambda^{(4)\!\dagger} U^{3\!\dagger} \Lambda^{(3)\!\dagger} U^{2\!\dagger} \Lambda^{(2)} U^{1\!\dagger} \Lambda^{(1)} \big).
\end{aligned}
\label{eq:trace_reduction}
\end{equation}

\paragraph*{Index expansion and Haar integration.}
Writing the two traces with explicit indices, e.g.
\begin{equation}
\mathrm{Tr}\!\big(\Lambda^{(1)\!\dagger} U^1 \Lambda^{(2)\!\dagger} U^2 \Lambda^{(3)} U^3 \Lambda^{(4)} U^4 \big)
=\Lambda^{(1)\!*}_{i_1}\, U^1_{i_1 i_2}\, \Lambda^{(2)\!*}_{i_2}\, U^2_{i_2 i_3}\, \Lambda^{(3)}_{i_3}\, U^3_{i_3 i_4}\, \Lambda^{(4)}_{i_4}\, U^4_{i_4 i_1},
\end{equation}
and similarly for the conjugate trace, the averages over $U^1, U^2, U^3$ factorize and can be performed with the second-moment Weingarten formula,
\begin{equation}
\int_{\mathrm{Haar}} dU\; U_{i\alpha} U^*_{j\beta} U_{k\gamma} U^*_{l\delta}
=\frac{\delta_{ij}\delta_{\alpha\beta}\delta_{kl}\delta_{\gamma\delta}+\delta_{il}\delta_{\alpha\delta}\delta_{kj}\delta_{\gamma\beta}}{d^2-1}
-\frac{\delta_{ij}\delta_{\alpha\delta}\delta_{kl}\delta_{\gamma\beta}+\delta_{il}\delta_{\alpha\beta}\delta_{kj}\delta_{\gamma\delta}}{d(d^2-1)}.
\label{eq:weingarten_2nd_moment}
\end{equation}
Applying Eq.~\eqref{eq:weingarten_2nd_moment} independently to $U^1$, $U^2$, and $U^3$, one organizes the resulting contractions into a $1/d$ expansion. The leading terms are either (i) fully “disconnected’’ (no phase correlations between segments), or (ii) products of “connected’’ pairings that are precisely the spectral form factors of the individual segments.

\paragraph*{Eigenvalue integrals and leading behavior.}
Introduce the (two-point) spectral form factor (SFF)
\begin{equation}
R_2(\tau)=\Big\langle \big|\mathrm{Tr}(e^{i H \tau})\big|^2 \Big\rangle_{H\sim\mathrm{GUE}}
=\Big\langle \sum_{a,b=1}^d e^{i(\lambda_a-\lambda_b)\tau}\Big\rangle,
\label{eq:R2_def}
\end{equation}
so that the leading large-$d$ contribution to Eq.~\eqref{eq:trace_reduction} reads
\begin{equation}
F^{(1)}(t;t_s)
=\frac{d^6}{(d^2-1)^3}
+\frac{d^8}{(d^2-1)^3}\!\left(\frac{R_2(t-t_s)}{d^2}\right)^{\!2}\!\left(\frac{R_2(t_s)}{d^2}\right)^{\!2}
+\mathcal{O}(\frac{1}{d})
=1+d^2\!\left(\frac{R_2(t-t_s)}{d^2}\right)^{\!2}\!\left(\frac{R_2(t_s)}{d^2}\right)^{\!2}
+\mathcal{O}(\frac{1}{d}).
\label{eq:F1_leading}
\end{equation}
Setting $t_s=t$ recovers the single-segment (no-quench) result. The eigenvalue measures $D\lambda_a$ are those induced by GUE; only the SFFs survive at leading order.

\paragraph*{GUE spectral form factor.}
For GUE (with the standard Wigner semicircle scaling), one may write
\begin{equation}
R_2(t)= d + d^2\, r_1(t)^2 + d\, r_2(t),
\qquad r_1(t)=\frac{J_1(2t)}{t},
\label{eq:R2_GUE}
\end{equation}
where $r_1(t)\sim t^{-3/2}$ for $t\gg 1$ and $r_2(t)$ encodes the connected “ramp’’ contribution. Equivalently,
\begin{equation}
\frac{R_2(t)}{d^2}= r_1(t)^2 + \mathcal{O}(\frac{t}{d^2}).
\end{equation}
The qualitative time dependence is
\begin{equation}
R_2(t)\sim
\begin{cases}
d^2, & t \ll t_{\mathrm{Th}},\\[2pt]
\sqrt{d}, & t \sim t_{\mathrm{Th}},\\[2pt]
d, & t \gtrsim t_p,
\end{cases}
\end{equation}
where $t_{\mathrm{Th}}$ is the Thouless time and $t_p$ the plateau time.

\paragraph*{Consequence for the frame potential.}
Inserting Eq.~\eqref{eq:R2_GUE} into Eq.~\eqref{eq:F1_leading} shows that, when the switch time satisfies $t_s\gtrsim t_{\mathrm{Th}}$, both $R_2(t_s)/d^2$ and $R_2(t-t_s)/d^2$ are bounded by $\mathcal{O}(1/d)$ as $d\to\infty$, hence the second term in Eq.~\eqref{eq:F1_leading} vanishes in the thermodynamic limit. Consequently, the frame potential attains the Haar value,
\begin{equation}
\lim_{d\to\infty} F^{(1)}(t;t_s)=1,
\end{equation}
i.e., the single-quench GUE evolution realizes a unitary 1-design.

\section{Appendix C: The single-quench evolution of integrable models}\label{appendix_integrable_model}
This appendix presents additional numerics for the frame potential (FP) under the single-quench evolution in \emph{integrable} models. We consider (i) the SYK$_2$ model, which is free and integrable, and (ii) the Richardson model, which is interacting yet integrable.
\subsection{SYK$_2$ model (free, integrable).}
The Hamiltonian of the complex SYK$_2$ model is
\begin{equation}
    \label{eqn:SYK2}
    H = i\sum_{ 1\leq j<k \leq N}J_{jk} c_j^{\dagger} c_k
\end{equation}
with $J_{ij}$ a Gaussian random with zero mean and satisfies $J_{ij}=J_{ji}^*$.
\begin{figure*}[tb]
  \centering
\includegraphics[width=0.98\textwidth]{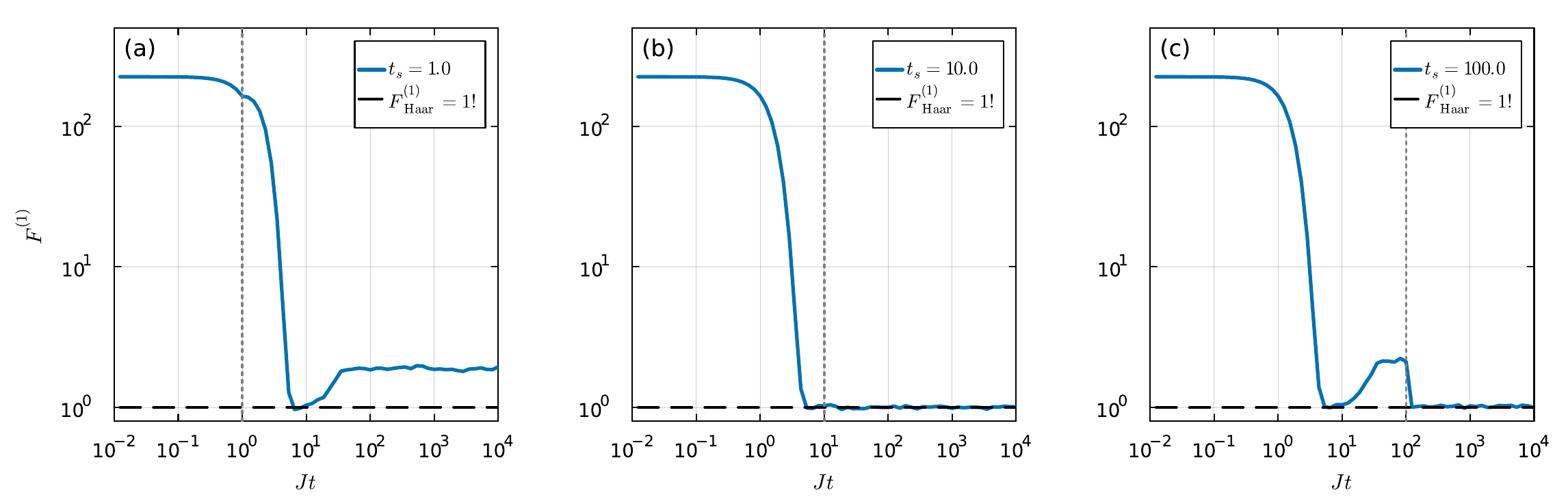}
  \caption{First-order frame potential $F^{(1)}$ versus total evolution time for single-quench dynamics of the complex-fermion SYK$_2$ model in a fixed charge sector. Parameters: $N=6$, $q=2$, $J=1$; averages over $10^4$ disorder realizations. The Haar-random benchmark $F^{(1)}=1$ is shown as a black dashed curve. The switch time $t_s$ is indicated by a gray dot vertical line in each panel: (a) $Jt_s=1.0$ (before the Thouless time $t_{\mathrm{Th}}$), (b) $Jt_s=10.0$ (near $t_{\mathrm{Th}}$), and (c) $Jt_s=100.0$ (after $t_{\mathrm{Th}}$). At late times, the plateau of $F^{(2)}$ lies above the Haar value for $t_s<t_{\mathrm{Th}}$, while it saturates the Haar value when $t_s\approx t_{\mathrm{Th}}$ or $t_s>t_{\mathrm{Th}}$.}
  \label{appendix_CSYK2_k1_fig}
\end{figure*}
\begin{figure*}[tb]
  \centering
\includegraphics[width=0.98\textwidth]{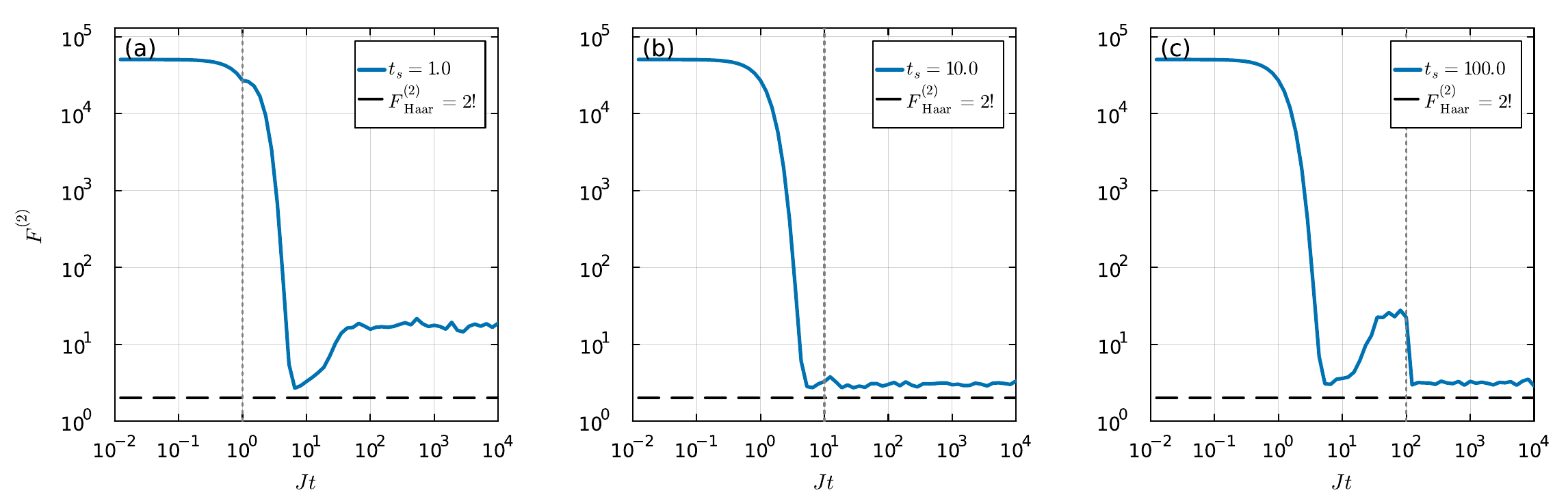}
  \caption{First-order frame potential $F^{(2)}$ versus total evolution time for single-quench dynamics of the complex-fermion SYK$_2$ model in a fixed charge sector. Parameters: $N=6$, $q=2$, $J=1$; averages over $10^4$ disorder realizations. The Haar-random benchmark $F^{(2)}=2$ is shown as a black dashed curve. The switch time $t_s$ is indicated by a gray dot vertical line in each panel: (a) $Jt_s=1.0$ (before the Thouless time $t_{\mathrm{Th}}$), (b) $Jt_s=10.0$ (near $t_{\mathrm{Th}}$), and (c) $Jt_s=100.0$ (after $t_{\mathrm{Th}}$). At late times, the plateau of $F^{(2)}$ lies above the Haar value for all different $t_s$.}
  \label{appendix_CSYK2_k2_fig}
\end{figure*}
Let $t_s$ denote the switch time and $t_{\mathrm{Th}}$ the Thouless time. For $t_s \ge t_{\mathrm{Th}}$, the first-order frame potential $F^{(1)}$ saturates the Haar value at late times, whereas higher-order frame potentials $F^{(k>1)}$ do not. As shown in Fig.~\ref{appendix_CSYK2_k1_fig}, when $t_s \gtrsim t_{\mathrm{Th}}$ the curve for $F^{(1)}$ reaches the Haar plateau [Fig.~\ref{appendix_CSYK2_k1_fig}(b) and (c)], indicating realization of a unitary 1-design; for $t_s < t_{\mathrm{Th}}$ the late-time plateau lies above the Haar value [Fig.~\ref{appendix_CSYK2_k1_fig} (a)], so a 1-design is not achieved. Moreover, Fig.~\ref{appendix_CSYK2_k2_fig} shows that $F^{(2)}$ in the complex SYK$_2$ model fails to reach its Haar value for any $t_s$ (before or after $t_{\mathrm{Th}}$). Instead, it saturates at the so-called Gaussian Haar value \cite{Tiutiakina2023FramePO}, which indicates that the single-quench model can still realize Brownian dynamics after the quench. More information on this can be found in SM \cite{SM}. Thus, under the single-quench protocol, the integrable SYK$_2$ ensemble realizes a unitary 1-design but no higher designs, in contrast to the chaotic SYK$_4$ case. This supports the achievable design order as an operational diagnostic of chaos.

\subsection{Richardson model (interacting, integrable).}
For the Richardson model with random single-particle energies $\{\epsilon_j\}$, the Hamiltonian is
\begin{equation}
    H = \frac{J}{N}\sum_{i,j=1}^N S^{+}_i S^{-}_j + \sum_{j=1}^N \epsilon_j S_j^z .
\end{equation}
There exists a complete set of mutually commuting Gaudin charges ${R_i({\epsilon})}$ such that $[R_i,R_j]=0$ and $[H,R_i]=0$; in fact, $H$ lies in the linear span of ${R_i}$. The model is therefore integrable—interacting, yet endowed with randomness through ${\epsilon_j}$. Here, the prefactor $1/N$ ensures that both terms in the Hamiltonian scale equally in the large-$N$ limit.

\begin{figure*}[tb]
  \centering  \includegraphics[width=0.98\textwidth]{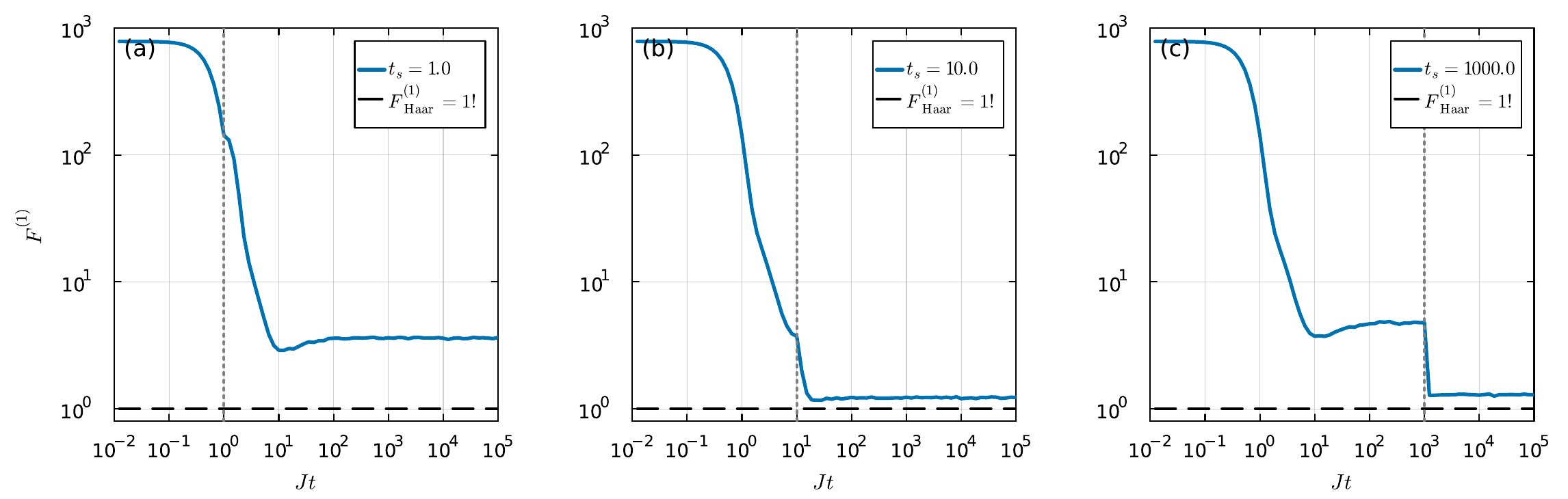}
  \caption{First-order frame potential $F^{(1)}$ versus total evolution time for single-quench dynamics of the Richardson model. Parameters: $N=8$, $q=2$, $J=1$; averages over $10^4$ disorder realizations. The random local energy $\{\epsilon_j \}$ is chosen to be Gaussian random with zero mean and variance $\sigma=1$. The Haar-random benchmark $F^{(1)}=1$ is shown as a black dashed curve. The switch time $t_s$ is indicated by a gray dashed vertical line in each panel: (a) $t_s=1.0$ (before the Thouless time $t_{\mathrm{Th}}$), (b) $t_s=10.0$ (near $t_{\mathrm{Th}}$), and (c) $t_s=1000.0$ (after $t_{\mathrm{Th}}$).}
  \label{appendix_random_spin_k1_fig}
\end{figure*}

The first order FP of the Richardson model as a function of total evolution time is illustrated in Fig.~\ref{appendix_random_spin_k1_fig},  and it shows that its $F^{(1)}$ fails to reach its Haar value for any $t_s$ (before or after $t_{\mathrm{Th}}$). Thus, under the single-quench protocol, this random integrable and interacting model fails to realize even a unitary 1-design. Compared with the complex SYK$_2$ model, which is integrable and free but able to realize a 1-design, this integrable but interacting model fails to realize a 1-design, which is interesting, and it could indicate that the SYK$_2$ model is more \textit{"random"} than the Richardson model, although it is non-interacting.

Across these integrable examples, the single-quench evolution fails to realize higher $(k\geq 2)$ unitary designs even when the switch time exceeds the Thouless time. This supports the view that the ability of the protocol to produce a unitary design is a diagnostic of \emph{chaos}: chaotic dynamics exhibit FP saturation at $t_s \gtrsim t_{\mathrm{Th}}$, whereas integrable dynamics do not.

\section{Appendix D: More numerical results about single-quench evolution of chaotic models}
In this appendix, we present additional numerical results for single-quench dynamics in chaotic models, including the complex SYK$_4$ model, the complex SYK$_4$ model with reduced-rank couplings, and the Majorana SYK$_4$ model.

\subsection{Complex SYK$_4$ model}
\begin{figure*}[tb]
  \centering
\includegraphics[width=0.98\textwidth]{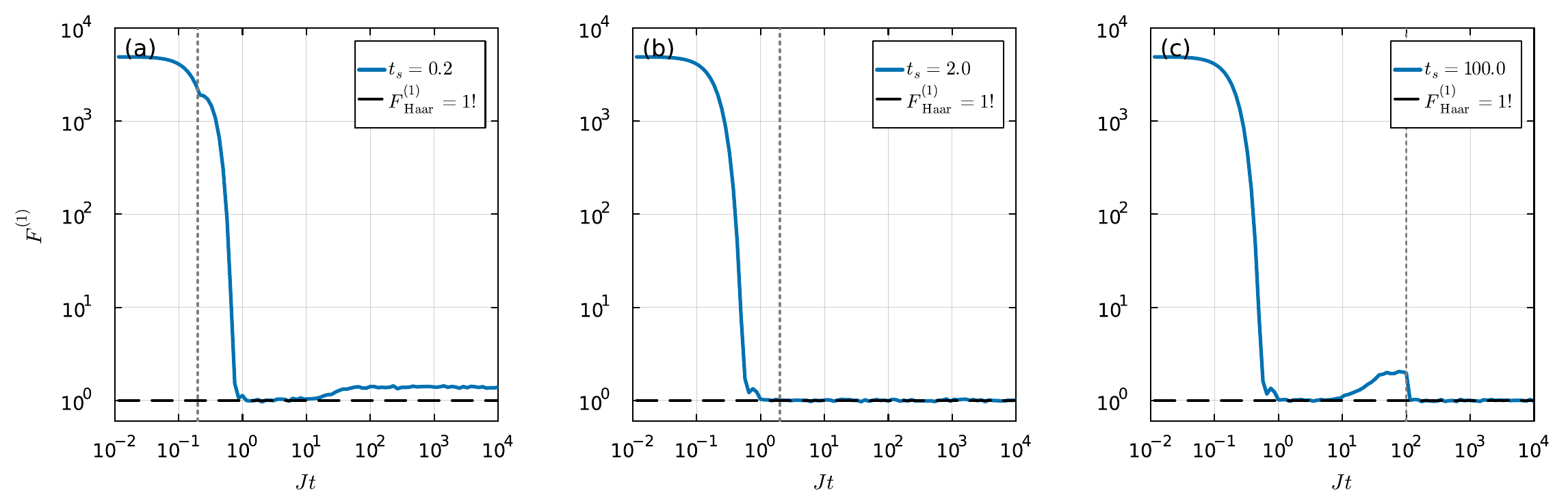}
  \caption{First-order frame potential $F^{(1)}$ versus total evolution time for single-quench dynamics of the complex-fermion SYK$_4$ model in a fixed charge sector. Parameters: $N=8$, $q=4$, $J=1$; averages over $4000$ disorder realizations. The Haar-random benchmark $F^{(1)}=1$ is shown as a black dashed curve. The switch time $t_s$ is indicated by a gray dashed vertical line in each panel: (a) $t_s=0.2$ (before the Thouless time $t_{\mathrm{Th}}$), (b) $t_s=2.0$ (near $t_{\mathrm{Th}}$), and (c) $t_s=100.0$ (after $t_{\mathrm{Th}}$). At late times, the plateau of $F^{(1)}$ lies above the Haar value for $t_s<t_{\mathrm{Th}}$, while it saturates the Haar value when $t_s\approx t_{\mathrm{Th}}$ or $t_s>t_{\mathrm{Th}}$.}
  \label{appendix_CSYK4_k1_fig}
\end{figure*}

\begin{figure*}[tb]
  \centering
\includegraphics[width=0.98\textwidth]{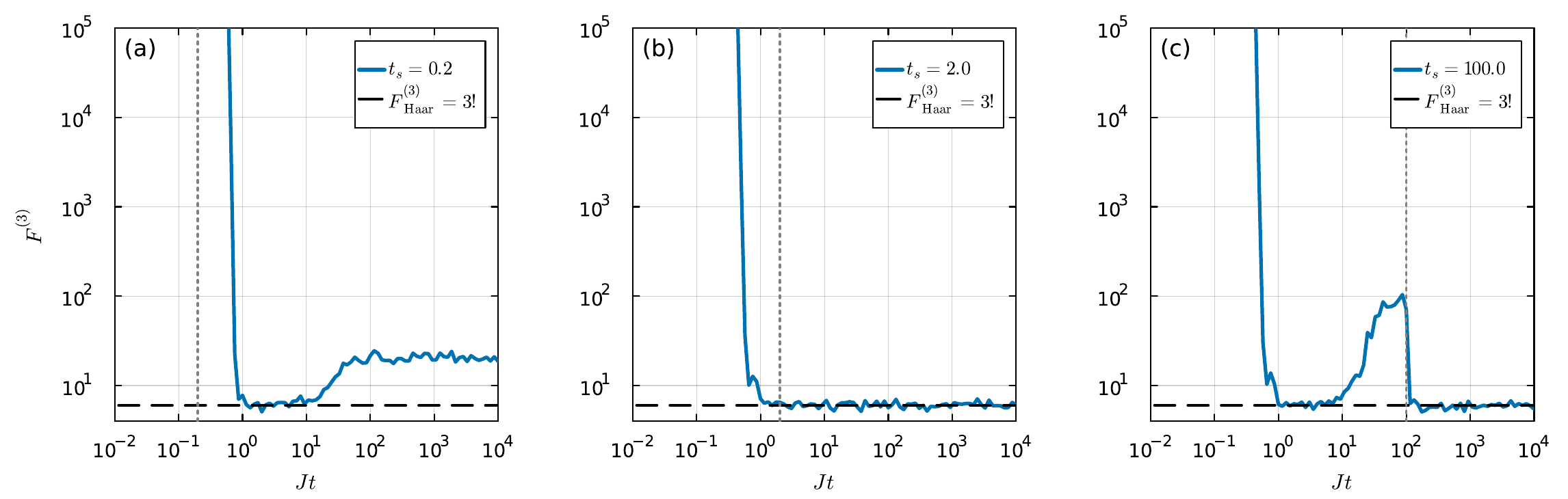}
  \caption{Third-order frame potential $F^{(3)}$ versus total evolution time for single-quench dynamics of the complex-fermion SYK$_4$ model in a fixed charge sector. Parameters: $N=8$, $q=4$, $J=1$; averages over $4000$ disorder realizations. The Haar-random benchmark $F^{(3)}=3!$ is shown as a black dashed curve. The switch time $t_s$ is indicated by a gray dot vertical line in each panel: (a) $Jt_s=0.2$ (before the Thouless time $t_{\mathrm{Th}}$), (b) $Jt_s=2.0$ (near $t_{\mathrm{Th}}$), and (c) $Jt_s=100.0$ (after $t_{\mathrm{Th}}$).}
  \label{appendix_CSYK4_k3_fig}
\end{figure*}

\begin{figure*}[tb]
  \centering
\includegraphics[width=0.98\textwidth]{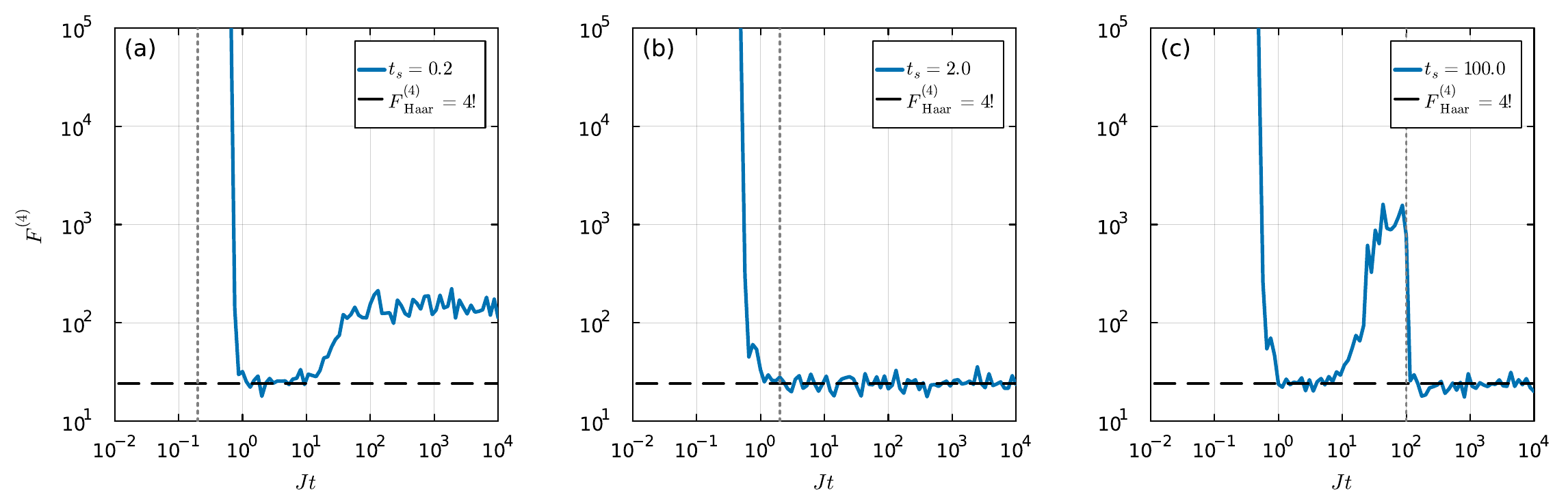}
  \caption{Fourth-order frame potential $F^{(4)}$ versus total evolution time for single-quench dynamics of the complex-fermion SYK$_4$ model in a fixed charge sector. Parameters: $N=8$, $q=4$, $J=1$; averages over $4000$ disorder realizations. The Haar-random benchmark $F^{(4)}=4!$ is shown as a black dashed curve. The switch time $t_s$ is indicated by a gray dot vertical line in each panel: (a) $Jt_s=0.2$ (before the Thouless time $t_{\mathrm{Th}}$), (b) $Jt_s=2.0$ (near $t_{\mathrm{Th}}$), and (c) $Jt_s=100.0$ (after $t_{\mathrm{Th}}$).}
  \label{appendix_CSYK4_k4_fig}
\end{figure*}

The first-, third-, and fourth-order FPs of the complex SYK$_4$ model are shown in Figs.~\ref{appendix_CSYK4_k1_fig}, \ref{appendix_CSYK4_k3_fig}, and \ref{appendix_CSYK4_k4_fig}, respectively. As evident there, the single-quench evolution attains the Haar value once the switch time satisfies $t_s \ge t_{\mathrm{Th}}$, indicating that this protocol realizes (at least) a unitary 4-design.

\begin{figure}[t] 
    \centering \includegraphics[width=0.46\textwidth]{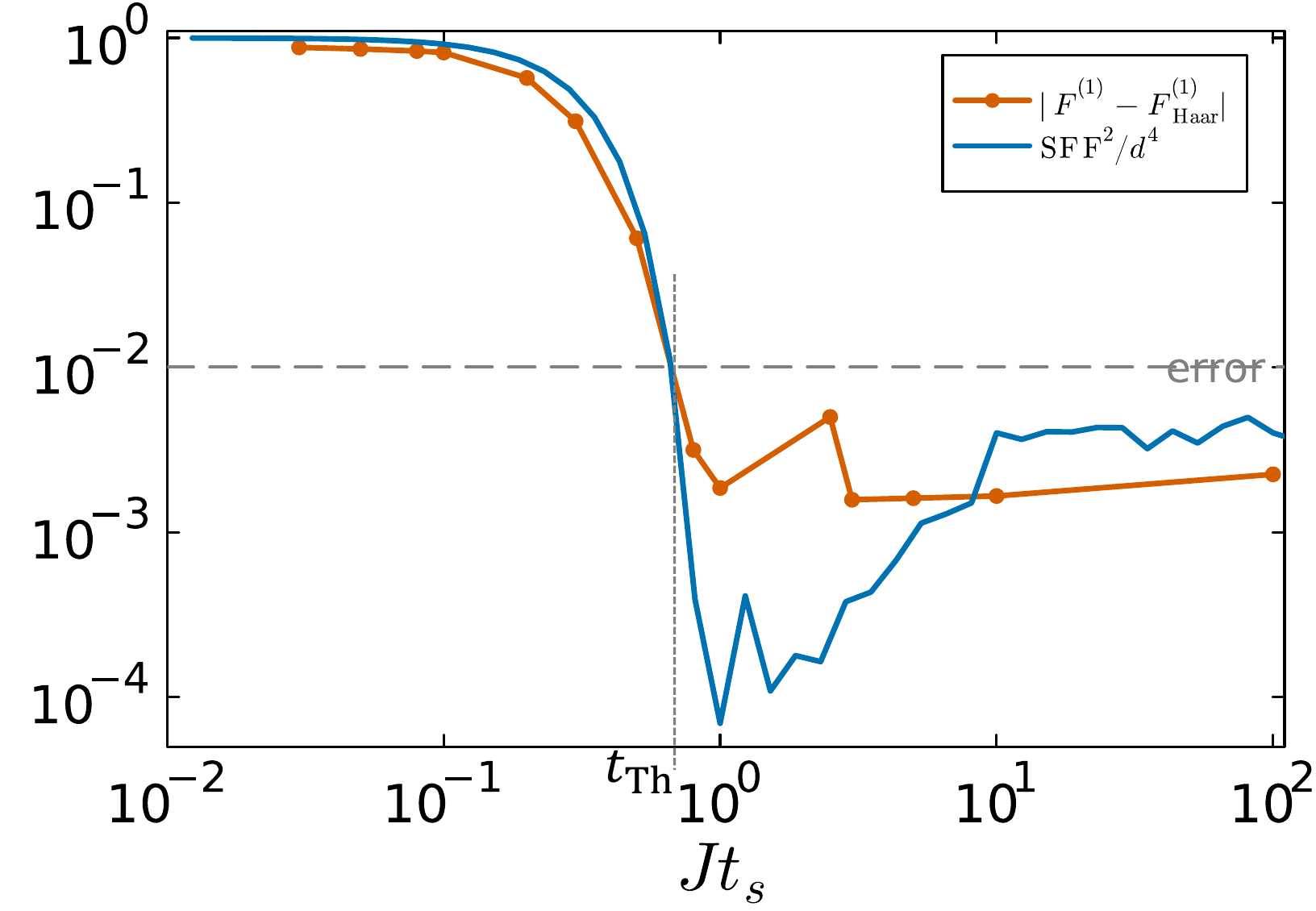} 
    \caption{Gap $\Delta_{\mathrm{FP}}(t_s)$ between the late-time plateau of the first-order frame potential $F^{(1)}$ and its Haar value for single-quench evolution of the complex SYK$_4$ model (fixed charge sector), compared to $\mathrm{SFF}^2$, versus switch time $t_s$. Parameters: $N=6$, charge $q=2$, $J=1$, averaged over $N_{\mathrm{rand}}=10^4$ disorder realizations. We define the Thouless time $t_{\mathrm{Th}}$ as the earliest $t_s$ for which $\Delta_{\mathrm{FP}}(t_s)$ is consistent with zero within statistical uncertainty (gray dashed line), estimated as $\sqrt{1/[\sqrt{N_{\mathrm{rand}}}(\sqrt{N_{\mathrm{rand}}}-1)]}\simeq 0.01$.
}
\label{appendix_CSYK4_crossover_k1_fig}
\end{figure}

\begin{figure}[t] 
    \centering \includegraphics[width=0.46\textwidth]{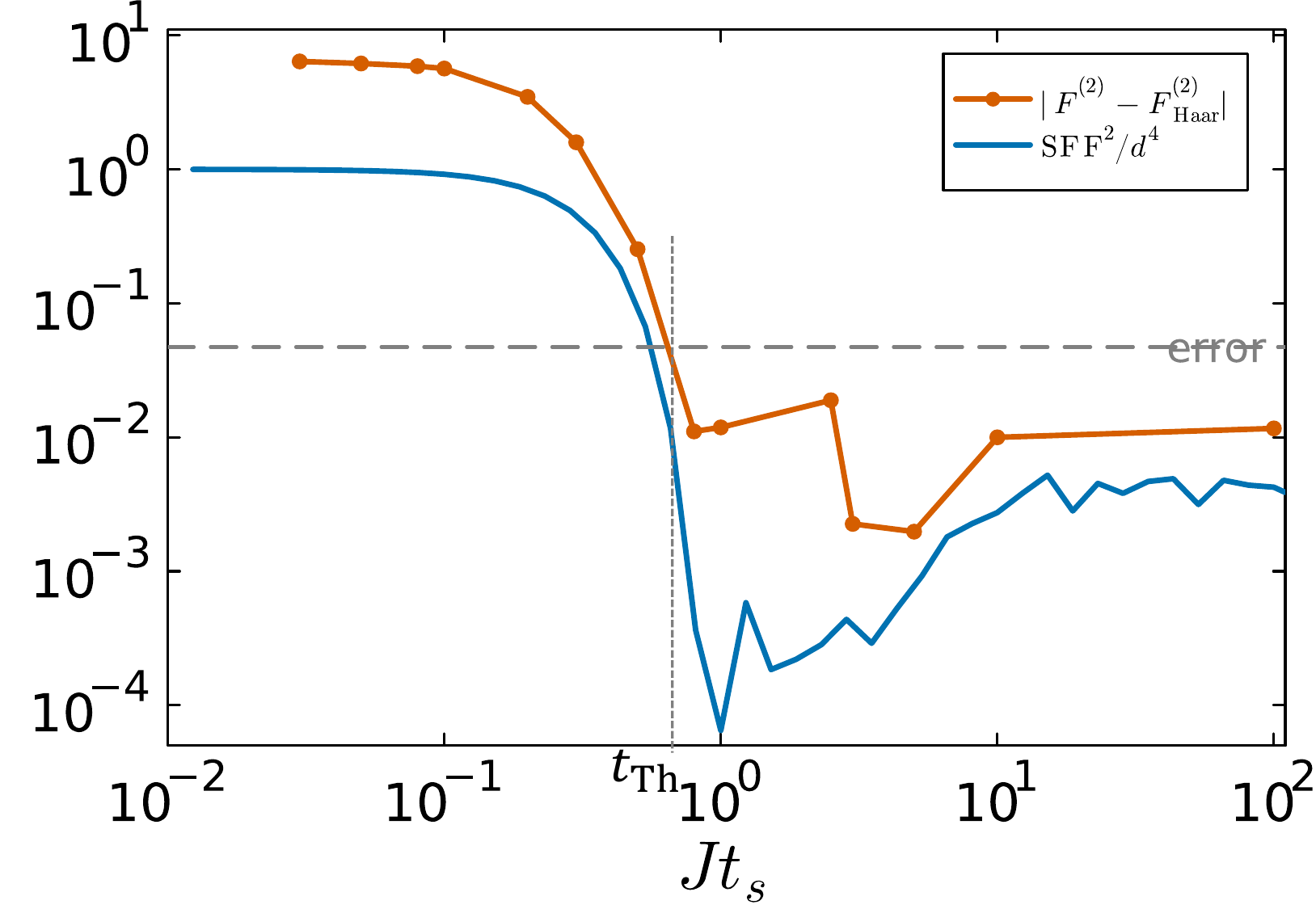} 
    \caption{Gap $\Delta_{\mathrm{FP}}(t_s)$ between the late-time plateau of the second-order frame potential $F^{(2)}$ and its Haar value for single-quench evolution of the complex SYK$_4$ model (fixed charge sector), compared to $\mathrm{SFF}^2$, versus switch time $t_s$. Parameters: $N=6$, charge $q=2$, $J=1$, averaged over $N_{\mathrm{rand}}=10^4$ disorder realizations. We define the Thouless time $t_{\mathrm{Th}}$ as the earliest $t_s$ for which $\Delta_{\mathrm{FP}}(t_s)$ is consistent with zero within statistical uncertainty (gray dashed line), estimated as $\sqrt{(4!-2!)/[\sqrt{N_{\mathrm{rand}}}(\sqrt{N_{\mathrm{rand}}}-1)]}\simeq 0.047$.
}
\label{appendix_CSYK4_crossover_k2_fig}
\end{figure}

Define the $k$-th order FP gap as
\begin{equation}
    \Delta_{\mathrm{FP}}^{(k)}(t_s)\equiv F^{(k)}(t_s\to \infty)-F^{(k)}_{\mathrm{Haar}},
 \qquad F^{(k)}_{\mathrm{Haar}}=k!
\end{equation}
Figs.~\ref{appendix_CSYK4_crossover_k1_fig} and \ref{appendix_CSYK4_crossover_k2_fig} show $\Delta_{\mathrm{FP}}^{(1)}(t_s)$ and $\Delta_{\mathrm{FP}}^{(2)}(t_s)$ versus the switch time $t_s$ for single-quench evolution of the complex SYK$_4$ model, compared with the square of the normalized SFF. 
The gap vanishes at the Thouless time, providing an operational definition of $t_{\mathrm{Th}}$. Consistent with Fig.~3 in the main text, the extracted $t_{\mathrm{Th}}$ is independent of the FP order $k$. Moreover, the FP-defined $t_{\mathrm{Th}}$ lies close to the onset of the SFF’s linear ramp, whose precise location is difficult to determine due to intrinsic oscillations near the crossover.

\subsection{Complex SYK$_4$ model with reduced-rank couplings}
\begin{figure*}[t]
  \centering  \includegraphics[width=0.98\textwidth]{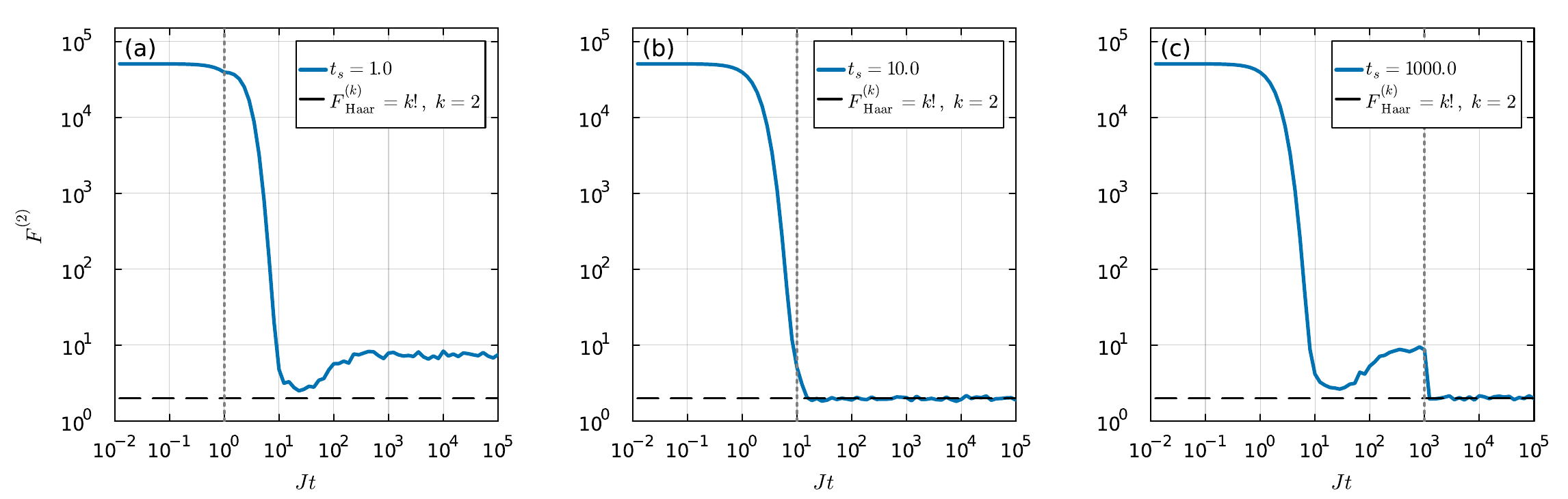}
  \caption{Second-order frame potential $F^{(2)}$ versus total evolution time for single-quench dynamics of the complex-fermion SYK$_4$ model with reduced rank couplings in a fixed charge sector. Parameters: $N=6$, $q=2$, $J=1$; averages over $10^4$ disorder realizations. The Haar-random benchmark $F^{(2)}=2$ is shown as a black dashed curve. The switch time $t_s$ is indicated by a gray dashed vertical line in each panel: (a) $t_s=1.0$ (before the Thouless time $t_{\mathrm{Th}}$), (b) $t_s=10.0$ (near $t_{\mathrm{Th}}$), and (c) $t_s=1000.0$ (after $t_{\mathrm{Th}}$). At late times, the plateau of $F^{(2)}$ lies above the Haar value for $t_s<t_{\mathrm{Th}}$, while it saturates the Haar value when $t_s\approx t_{\mathrm{Th}}$ or $t_s>t_{\mathrm{Th}}$.}
  \label{appendix_reduced_SYK4_fig}
\end{figure*}

We also consider the SYK model with reduced rank coupling whose Hamiltonian is given by \cite{baumgartner2024quantumsimulationsachdevyekitaevmodel}
\begin{equation}
    H = \sum_{ 1\leq i<j<k<l \leq N}J_{ijkl}^{\mathrm{red}} c_i^{\dagger} c_j^{\dagger} c_k c_l, 
\end{equation}
and $J_{ijkl}^{\mathrm{red}} = J_{ik}J_{jl}- J_{il}J_{jk}$, $J_{ij}$ is Gaussian with zero mean and satisfies $J_{ij}=J_{ji}^*$. This model is more friendly to experimental realization of the SYK model \cite{uhrich2023cavityquantumelectrodynamicsimplementation}. Its single-quench time evolution can also satisfy unitary $2$-designs, as shown in Fig.~\ref{appendix_reduced_SYK4_fig}.

\subsection{Majorana SYK$_4$ model}
\begin{figure*}[t]
  \centering  \includegraphics[width=0.98\textwidth]{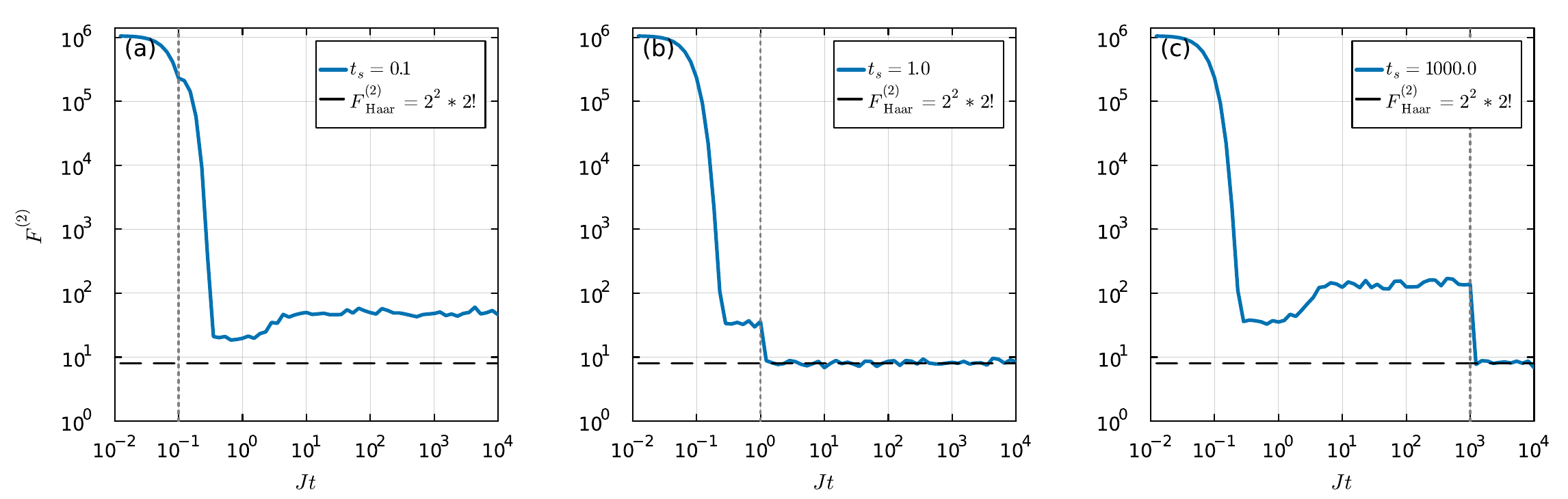}
  \caption{Second-order frame potential $F^{(2)}$ versus total evolution time for single-quench dynamics of the Majorana SYK$_4$ model. Parameters: $N=5$, $J=1$; averages over $10^3$ disorder realizations. The Haar-random benchmark $F^{(k)}=2^kk!$ for $k=2$ is shown as a black dashed curve. The switch time $t_s$ is indicated by a gray dashed vertical line in each panel: (a) $t_s=0.1$ (before the Thouless time $t_{\mathrm{Th}}$), (b) $t_s=1.0$ (near $t_{\mathrm{Th}}$), and (c) $t_s=1000.0$ (after $t_{\mathrm{Th}}$). At late times, the plateau of $F^{(2)}$ lies above the Haar value for $t_s<t_{\mathrm{Th}}$, while it saturates the Haar value (up to the symmetry factor $2^k$) when $t_s\approx t_{\mathrm{Th}}$ or $t_s>t_{\mathrm{Th}}$.}
  \label{appendix_Major_SYK4_fig}
\end{figure*}

The Hamiltonian of the Majorana SYK$_4$ model is
\begin{equation*}
    H=\sum_{i,j,k,l=1}^N J_{ijkl}\chi_i\chi_j\chi_k\chi_l
\end{equation*}
with Majorana fermions $\{\chi_i,\chi_j\}=2\delta_{ij}$. Its single-quench time evolution saturates the plateau value $2^k k!$, as shown in Fig.~\ref{appendix_Major_SYK4_fig}. The additional factor $2^k$ relative to the Haar value arises from the Majorana symmetry factor.

\section{Appendix E: Single-Quench vs. Brownian SYK}
\label{appendix_Analytic_Brownian}
\begin{figure}[t] 
    \centering \includegraphics[width=0.52\textwidth]{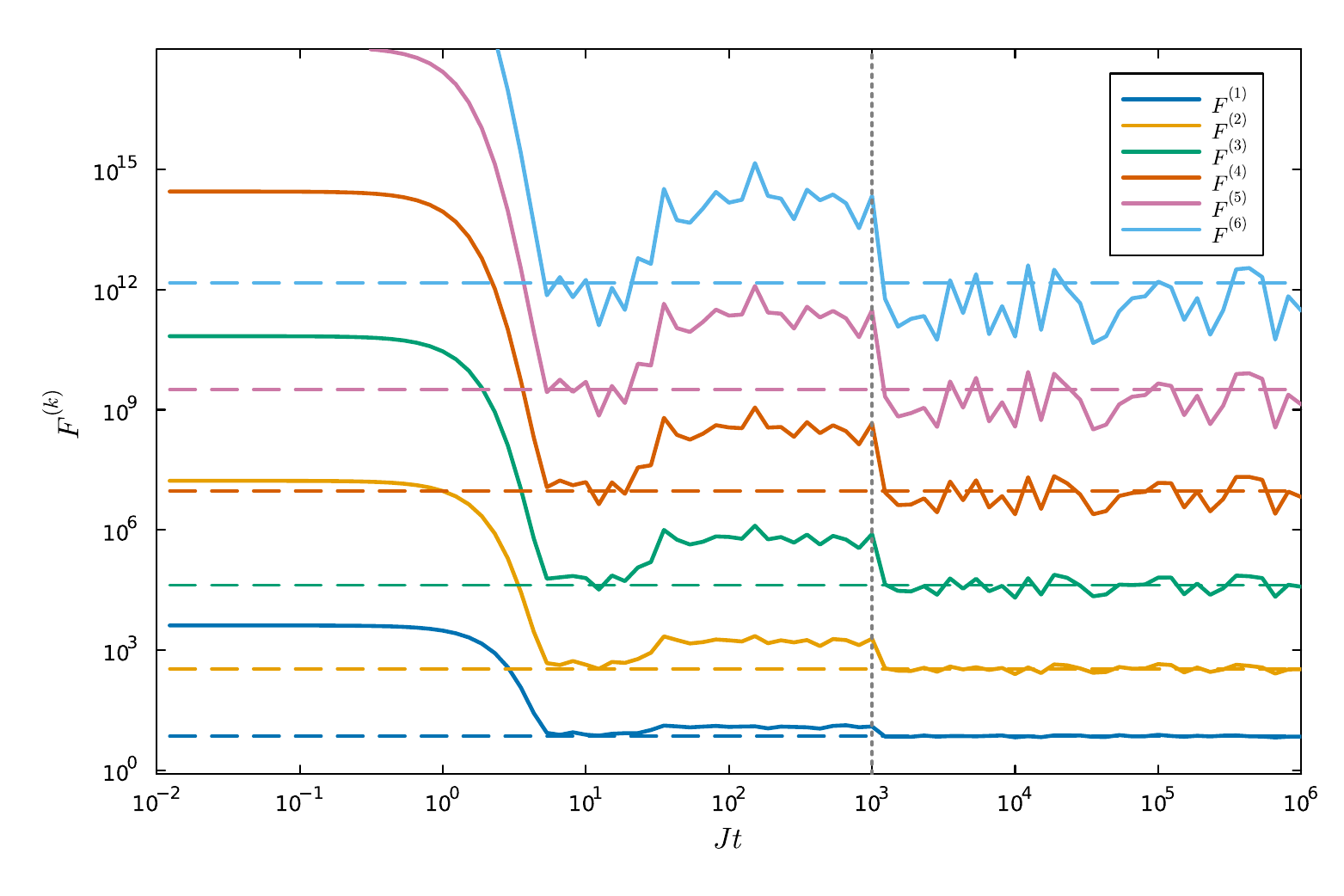} 
    \caption{Frame potential of complex SYK$_2$ for \textit{all} charged sectors and for $k=1$ to $k=6$ with $N=6$ fermions ($3000$ realizations). The colored dashed lines show the Gaussian Haar value (see eqn. \ref{eqn:GaussianHaarValue}). The grey dashed line indicates the switch time. Although the Gaussian Haar value becomes the Brownian SYK$_2$ value only for large $N$, it is still a good approximation at low $N$ and for low $k$. It can be seen in the plots that it is also the drop-off value of the single-quench model after the quench. Only for $k=6$, we can see a clear deviation, which is due to the low $N$ regime we are working in. We can therefore conclude that the single-quench model effectively becomes the Brownian model.}
\label{SYK2_brownian_comparison_fig}
\end{figure}

In this section, we give a short overview of some analytic results for complex Brownian SYK that have previously been studied in \cite{Jian2022LinearGO, Tiutiakina2023FramePO}. They can be used to benchmark the long-time behavior of the single-quench evolution, as our numerical data imply agreement with the Brownian case. For complex SYK$_4$, this agreement is expected, as both the Brownian evolution and our single-quench model saturate the Haar value k!. For SYK$_2$, however, this is a stronger result. Whether a single quench reproduces the Brownian average in general — e.g., for a larger class of integrable models that do not scramble to the Haar value — is a question we postpone for future work. \\
For the Brownian model, we can consider again the Hamiltonians of $SYK_4$ model
\begin{equation}
    H_{\text{SYK}} = \sum_{ 1\leq i<j<k<l \leq N}J_{ijkl} c_i^{\dagger} c_j^{\dagger} c_k c_l,
\end{equation}
and the Hamiltonion of $SYK_2$ model defined in Eq.\eqref{eqn:SYK2} but with time dependent couplings $J_{ijkl}(t)$ and $J_{ij}(t)$ s.t.
\begin{equation*} 
	\langle J_{ijkl}(t) J_{i^{'}j^{'}k^{'}l^{'}}(t') \rangle=\delta_{i,i^{'}}\delta_{j,j^{'}}\delta_{k,k^{'}}\delta_{l,l^{'}}\delta(t-t')\frac{3!J^2}{N^{3}} \qquad \text{and} \qquad \langle J_{ij}(t) J_{i^{'}j^{'}}(t') \rangle = \delta_{i,i'}\delta_{j,j'}\delta(t-t')\frac{J^2}{N}.
\end{equation*}
This model allows a convenient calculation of the Frame potential in terms of a path integral. We just need to notice that the forward-backward evolution $U^\dagger(t)V(t)$ can be effectively written as forward-only evolution $U(2t)$ because the couplings are gaussian variables centered around zero, rendering the sign irrelevant. \\
The calculation proceeds by summing the contributions from the classical saddle-points of the path integral. For SYK$_4$, this task is straightforward (see \cite{Jian2022LinearGO} for Majorana SYK and \cite{Tiutiakina2023FramePO} for the complex case). For the late time evolution they indeed find \footnote{Note that in \cite{Tiutiakina2023FramePO}, all charge sectors are summed up, so there is an additional factor of $N^k$. We only consider one fixed charge sector.}
\begin{equation*}
    F^{(k)}_{\text{cSYK}_4}(T\rightarrow\infty) = k!
\end{equation*}
which confirms that Brownian evolution saturates the Haar value. For SYK$_2$, the situation is more complicated. The problem is that the effective action appearing in the path integral exhibits a continuous $U(k)\times U(k)$ symmetry that is broken down to $SU(k)$ by the solution, which leads to additional Goldstone modes that activate higher order modes in the saddle point evaluation. In \cite{Tiutiakina2023FramePO} et al., was found that the long time behavior of the Frame potential scales as
\begin{equation*}
    F^{(k)}_{\text{SYK}_2}(T\rightarrow\infty) \sim N\cdot N^{k^2-1}
\end{equation*}
where $k^2-1$ is the number of massless Goldstone modes. Note that the Frame potential has been taken over \textit{all} charged sectors here, which gives the additional factor of $N$. \\
For large-N, this approaches the Frame potential of so-called Gaussian Haar Unitaries. That means, instead of a general $U$, we take $U=e^{iJ_{ij}c_i^\dagger c_j}$ where $J_{ij}$ is a Gaussian-valued hermitian matrix and we average over $J_{ij}$. The result reads \cite{Tiutiakina2023FramePO}
\begin{equation}
    \label{eqn:GaussianHaarValue}
    F^k_{\text{Gauss}} = \prod_{n=0}^{N-1}\frac{\Gamma(n+2k+1)\Gamma(n+1)}{\Gamma(n+k+1)^2}.
\end{equation}
For large $N$, the relative Frame potential $F^{(k)}(T)/F^{(k)}(0)$ with $F^{(k)}(0)=2^{Nk}$ can be approximated by
\begin{equation*}
    \lim_{N\to\infty}\Big(\frac{1}{N}\log\Big(2^{-Nk}F^{(k)}_{\text{Gauss}}\Big)\Big) \sim -k\log(2)+k^2\frac{\log(N)}{N} \sim \lim_{N\to\infty}\Big(\frac{1}{N}\log\Big(2^{-Nk}F^{(k)}_{\text{SYK}_2}(T\rightarrow\infty)\Big)\Big)
\end{equation*}
This approximation also holds well for small N and k, as shown in Fig.~\ref{SYK2_brownian_comparison_fig}.

\section{Appendix F: Statistical error bar for the $k$-th frame potential (finite sample)}\label{sec:fp_errorbar}
In this appendix, we provide the error-bar calculation for the finite-sample estimate of the $k$th FP used in the main-text Fig 3. Let $U_1,\dots,U_M$ be $M$ independent Haar-random unitaries in $U(d)$. The $k$-th frame potential is
\begin{equation}
\mathrm{FP}_k \;=\; \mathbb{E}_{U,V\sim \mathrm{Haar}}\!\left[\,\big|\Tr(U^\dagger V)\big|^{2k}\right].
\end{equation}
Given a finite sample $\{U_i\}_{i=1}^M$, we estimate $F^{(k)}$ using the off-diagonal (unbiased) $U$-statistic
\begin{equation}
F^{(k)}
\;=\;
\frac{1}{M(M-1)}\sum_{i\neq j}\big|\Tr(U_i^\dagger U_j)\big|^{2k}
\;=\;
\frac{2}{M(M-1)}\sum_{1\le i<j\le M} h_{ij},
\qquad
h_{ij}:=\big|\Tr(U_i^\dagger U_j)\big|^{2k}.
\label{eq:FPk_estimator}
\end{equation}
For Haar draws, each $W_{ij}:=U_i^\dagger U_j$ is itself Haar-random, hence the $h_{ij}$ are identically distributed and
\begin{equation}
\mathbb{E}[h_{ij}] = \mathbb{E}_{W\sim\mathrm{Haar}}\!\left[|\Tr(W)|^{2k}\right] = \mathrm{FP}_k.
\end{equation}

\paragraph{Variance and standard error.}
A key simplification here is that the kernel is \emph{degenerate} in the Hoeffding sense:
for any fixed $U_i$, the random unitary $U_i^\dagger U_j$ is Haar-distributed and independent of $U_i$, so
$\mathbb{E}[h_{ij}\,|\,U_i]=\mathrm{FP}_k$ is constant. Therefore, the usual $O(1/M)$ contribution to the variance
vanishes, and the leading variance is $O(1/M^2)$. A convenient leading-order error bar is
\begin{equation}
\mathrm{Var}\!\left(F^{(k)}\right)
\;\approx\;
\frac{2}{M(M-1)}\,\mathrm{Var}\!\left(|\Tr(W)|^{2k}\right),
\qquad W\sim \mathrm{Haar},
\label{eq:FPk_var_general}
\end{equation}
so the standard error scales as
\begin{equation}
\mathrm{SE}\!\left(F^{(k)}\right)
\;\approx\;
\sqrt{\frac{2}{M(M-1)}}\;
\sqrt{\mathrm{Var}\!\left(|\Tr(W)|^{2k}\right)}
\;=\;O(M^{-1}).
\label{eq:FPk_SE_scaling}
\end{equation}

\paragraph{Relating to the convention $N_{\mathrm{rand}}=M^2$.}
In the main text we report the sampling budget as
\begin{equation}
N_{\mathrm{rand}}=M^2,
\end{equation}
while the estimator \eqref{eq:FPk_estimator} uses $M(M-1)$ off-diagonal pairs. Since $M(M-1)=M^2\,(1-1/M)$, one may
substitute $M=\sqrt{N_{\mathrm{rand}}}$ to obtain
\begin{equation}
\mathrm{SE}\!\left(F^{(k)}\right)
\;\approx\;
\sqrt{\frac{2}{\sqrt{N_{\mathrm{rand}}}\,(\sqrt{N_{\mathrm{rand}}}-1)}}\;
\sqrt{\mathrm{Var}\!\left(|\Tr(W)|^{2k}\right)}
\;=\;O\!\left(N_{\mathrm{rand}}^{-1/2}\right).
\label{eq:FPk_SE_scaling_Nrand}
\end{equation}

\paragraph{$k=1$ example.}
In the small-moment regime (e.g.\ $2k\le d$), Haar moments give
\begin{equation}
\mathbb{E}\!\left[|\Tr(W)|^{2k}\right]=k!,
\qquad
\mathbb{E}\!\left[|\Tr(W)|^{4k}\right]=(2k)!,
\end{equation}
hence
\begin{equation}
\mathrm{Var}\!\left(|\Tr(W)|^{2k}\right)=(2k)!-(k!)^2.
\end{equation}
For $k=1$,
\begin{equation}
\mathrm{Var}\!\left(|\Tr(W)|^{2}\right)=2!-(1!)^2=1,
\end{equation}
so the standard error of the off-diagonal estimator is
\begin{equation}
\mathrm{SE}\!\left(F^{(1)}\right)
\;\approx\;
\sqrt{\frac{2}{M(M-1)}}.
\label{eq:SE_FP1}
\end{equation}
For example, with $M=100$ (so $N_{\mathrm{rand}}=M^2=10^4$),
\begin{equation}
\mathrm{SE}\!\left(F^{(1)}\right)\approx \sqrt{\frac{2}{100\cdot 99}}\simeq 1.4\times 10^{-2},
\end{equation}
so deviations at the $10^{-2}$ level are consistent with finite-sample fluctuations.


\twocolumngrid
\bibliography{ref.bib}

\end{document}